\documentclass[aps,prd,amsmath,amsfonts,amssymb,eqsecnum,nofootinbib,floatfix,secnumarabic,preprintnumbers,superscriptaddress,nobalancelastpage,twocolumn]{revtex4}
\usepackage[utf8]{inputenc}
\usepackage{color,graphicx,multirow,slashed,hyperref}
\usepackage[export]{adjustbox}

\usepackage{amssymb}
\usepackage{enumerate}
\usepackage{amssymb}
\usepackage{amsmath}
\usepackage{tikz}
\usepackage{feynmf}
\usepackage{makecell}
\usepackage{slashed}
\usepackage{natbib}
\usepackage{graphicx}
\usepackage{setspace}
\usepackage{tensor}
\usepackage{physics}
\usepackage{slashed}
\usepackage{pifont}

\hypersetup{colorlinks=true}

\makeatletter
\newcommand{\fmslash}[2][0mu]{%
  \mathchoice
    {\fmsl@sh\displaystyle{#1}{#2}}%
    {\fmsl@sh\textstyle{#1}{#2}}%
    {\fmsl@sh\scriptstyle{#1}{#2}}%
    {\fmsl@sh\scriptscriptstyle{#1}{#2}}}
\newcommand{\fmsl@sh}[3]{%
  \m@th\ooalign{$\hfil#1\mkern#2/\hfil$\crcr$#1#3$}}
\makeatother

\newcommand{\bea}{\begin{eqnarray}}
\newcommand{\eea}{\end{eqnarray} }
\newcommand{\amc}{{\sc MadGraph5\textunderscore}a{\sc MC@NLO}}

\newcommand{\mptvec}{{\vec{\fmslash P}_T}}

\definecolor{bananayellow}{rgb}{1.0, 0.88, 0.21}
\definecolor{asparagus}{rgb}{0.53, 0.66, 0.42}

\begin{document}

\date{\today}

\title{Resolving Combinatorial Ambiguities in Dilepton $t \bar t$ Event Topologies with Neural Networks}

\author{Haider Alhazmi} 
\email{haider@ku.edu}
\affiliation{Department of Physics and Astronomy, University of Kansas, Lawrence, KS 66045, USA}
\affiliation{Department of Physics, Jazan University, Jazan 45142, Saudi Arabia}

\author{Zhongtian Dong} 
\email{cdong@ku.edu}
\affiliation{Department of Physics and Astronomy, University of Kansas, Lawrence, KS 66045, USA}

\author{Li Huang}
\email{huangli@ucas.ac.cn}
\affiliation{International Centre for Theoretical Physics Asia-Pacific, University of Chinese Academy of Sciences, 100190 Beijing, China}
\affiliation{Taiji Laboratory for Gravitational Wave Universe, University of Chinese Academy of Sciences, 100049 Beijing, China}

\author{\\ Jeong Han Kim}
\email{jeonghan.kim@cbu.ac.kr}
\affiliation{Department of Physics, Chungbuk National University, Cheongju, Chungbuk 28644, Korea}
\affiliation{Center for Theoretical Physics of the Universe, Institute for Basic Science, Daejeon 34126, Korea}
\affiliation{School of Physics, KIAS, Seoul 02455, South Korea}

\author{Kyoungchul Kong}
\email{kckong@ku.edu}
\affiliation{Department of Physics and Astronomy, University of Kansas, Lawrence, KS 66045, USA}

\author{David Shih}
\email{shih@physics.rutgers.edu}
\affiliation{NHETC, Dept. of Physics and Astronomy, Rutgers, The State University of NJ Piscataway, NJ 08854, USA}

\begin{abstract}
We study the potential of deep learning to resolve the combinatorial problem in SUSY-like events with two invisible particles at the LHC. As a concrete example, we focus on dileptonic $t \bar t$ events, where the combinatorial problem becomes an issue of binary classification: pairing the correct lepton with each $b$ quark coming from the decays of the tops. We investigate the performance of a number of machine learning algorithms, including attention-based networks, which have been used for a similar problem in the fully-hadronic channel of $t\bar t$ production; and the Lorentz Boost Network, which is motivated by physics principles. We then consider the general case when the underlying mass spectrum is unknown, and hence no kinematic endpoint information is available. Compared against existing methods based on kinematic variables, we demonstrate that the efficiency for selecting the correct pairing is greatly improved by utilizing deep learning techniques.
\end{abstract}


\maketitle

\section{Introduction}
\label{sec:intro}

Signatures with missing transverse momentum ($\mptvec$) are one of the most exciting classes of events at the Large Hadron Collider (LHC) and future colliders. They are produced by well-motivated scenarios of physics beyond the Standard Model (BSM), including supersymmetry and dark matter.
Unfortunately, events with $\mptvec$ are difficult to interpret and analyze due to {\it instrumental effects}, {\it unknown nature of the invisible particles}, and {\it incomplete kinematic information} \cite{Debnath:2017ktz}.  

One common approach to analyzing $\mptvec$ events is to hypothesize a certain event topology, and design suitable event variables adapted to this interpretation \cite{Barr:2011xt}. Already at this stage, one faces a combinatorial problem: {\it how to associate the reconstructed objects in the event with the elementary particles in the final state of the event topology}. The most common practice in resolving the combinatorial problem is to choose the ``best" assignment event by event. In this case one tries to design an algorithm (typically involving kinematic variables) which will single out one (or maybe several) among many possible assignments as the most likely ``correct" assignment. Then the value of the kinematic variable obtained with this specific choice is used for further analysis. 
In the presence of $\mptvec$, the combinatorics problem becomes more severe due to the unknown momenta of the invisible particles. 
To address the combinatorial problem properly, various strategies have been proposed, depending on the length of cascade decays. We refer to Ref. \cite{Debnath:2017ktz} and references therein for existing methods. 

In this paper, we study the combinatorial problem in events with $\mptvec$ using supervised machine learning (ML). As a concrete example, we will consider  dileptonic $t\bar t$ production. In this case, the combinatorial ambiguity is simply two-fold: how to correctly pair the two $b$-quarks with the two leptons in every event. In one sense, the two-fold ambiguity of dileptonic $t\bar t$ is the simplest combinatorial problem; yet on the other hand, the presence of two missing neutrinos brings additional challenges in reconstructing the final state. In any event, the two-fold ambiguity can be mapped to a straightforward binary classification task in supervised ML.

Supervised ML methods have recently been applied with much success to many areas of high energy physics (HEP), including jet and event classification \cite{Feickert:2021ajf}. ML methods -- especially those based on deep neural networks -- are able to learn subtle correlations in a high-dimensional space, and so often outperform more conventional methods based on physics-motivated high-level features (i.e.\ specially-designed kinematic variables). However, despite all of this recent activity, so far the combinatorial problem has not received much attention. Only very recently have there been studies of particular ML methods -- those utilizing the permutation-invariant structure of attention-based neural networks --  for the combinatorial problem in fully hadronic $t\bar t$ production \cite{Shmakov:2021qdz,Fenton:2020woz,Lee:2020qil}.
Also, other final states such as $t \bar t h$ \cite{Shmakov:2021qdz}, four tops \cite{Shmakov:2021qdz,Kim:2021wrr}, $HZ$ \cite{Kim:2021wrr}, and
stop pair with RPV \cite{Badea:2022dzb} have been studied in the fully hadronic channel using ML methods.

There are already many existing, non-ML strategies developed for the dilepton channel, such as endpoint methods \cite{Rajaraman:2010hy,Barr:2011xt,Baringer:2011nh,Choi:2011ys,Debnath:2017ktz}, hemisphere method \cite{Ball:2007zza,Matsumoto:2006ws,Cho:2007dh,Nojiri:2008hy}, topness \cite{Graesser:2012qy}, kinematic likelihood fit \cite{Erdmann:2013rxa} etc. Here we will compare a number of new ML-based methods, including the attention-based methods explored in  \cite{Shmakov:2021qdz,Fenton:2020woz,Lee:2020qil}, and benchmark them against the existing methods.   We will demonstrate that the new ML approaches lead to significantly improved ability in resolving the two-fold ambiguity compared to previous methods. More accurate solutions to the combinatorial problem of dileptonic $t\bar t$ could have many potential applications, including:
\begin{enumerate}
    \item Resolving the two-fold ambiguity in dilepton top quark pair production is of high importance for precision measurements of top quark and Higgs properties. For example, Refs. \cite{Goncalves:2018agy,Goncalves:2021dcu,Barman:2021yfh} aim to measure the Yukawa coupling and the CP-phase in the Top-Higgs interaction via $t\bar t h$ production in the dilepton channel, utilizing the kinematic methods proposed in Refs. \cite{Debnath:2017ktz,Jackson:2017gcy}. 
    
    \item Another non-trivial example is double Higgs production in the $h h \to (b\bar b) (WW^\ast) \to (b\bar b) (\ell^+\ell^- \nu \bar \nu)$ final state, where the dominant background is the dilepton $t\bar t$ production. Recent studies  \cite{Kim:2018cxf,Kim:2019wns} adopt a traditional $\chi^2$ method (Topness and Higgsness) to enhance the signal sensitivity of double Higgs production. The method attempts to solve the combinatorial problem by choosing the smallest $\chi^2$ value of all possible combinations. We expect that hybrid methods with ML would resolve the two-fold ambiguity better, eventually leading to the improved signal sensitivity.   
     
    \item The dilepton channel resembles signatures arising in various BSM scenarios, where the two missing particles could be dark matter (DM) candidates. Therefore our results will be valuable in reducing the $t\bar t$ backgrounds in the search for any new physics in the same final state. Also, we will generalize our method to arbitrary mass spectra, in order to address the two-fold ambiguity in the new physics scenarios. 

\end{enumerate}

This paper is structured as follows. We begin our discussion by describing the event generation and setup of ML methods in section \ref{sec:setup}. Our investigation of the performance of various machine learning algorithms is contained in section \ref{sec:ML}. 
In subsection \ref{sec:comparison}, we summarize our findings and compare various ML methods against one another as well as against existing approaches (which are briefly described in Appendix \ref{sec:existingmethods}).
In section \ref{sec:bsm} we study the combinatorial problem without the prior knowledge of the mass information, considering the $t \bar t$-like event topology. Section \ref{sec:conclusion} is reserved for discussion.

\section{Setup}
\label{sec:setup}

\subsection{Details of the simulation}
\label{sec:data}

\begin{figure}[t]
\begin{center}
\includegraphics[width=0.47\textwidth]{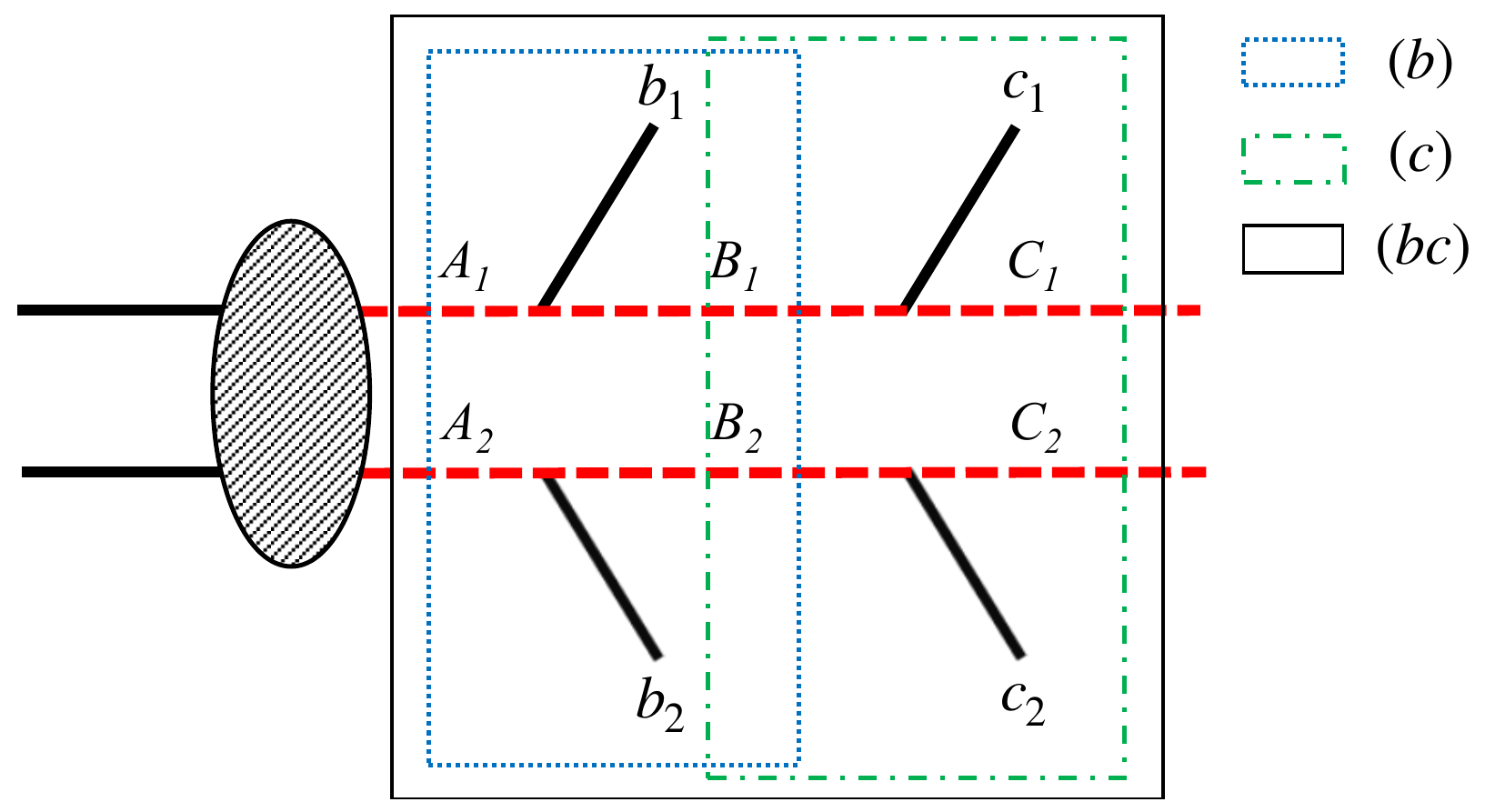}
\caption{The event topology considered in this paper, together with the three possible subsystems. The blue dotted, the green dot-dashed, and the black solid boxes indicate the subsystems $(b)$, $(c)$, and $(bc)$, respectively. The figure is taken from Ref. \cite{Cho:2014naa}.}
\label{fig:diagram}
\end{center}  
\end{figure}
The event topology considered in this article is depicted in Fig. \ref{fig:diagram}, where $A_i$,$B_i$ and $C_i$ ($i=1,2$ to denote two sides of decays) are top quarks ($t$), $W$-bosons and neutrinos ($\nu$), respectively. (In section \ref{sec:bsm}, we will assume they are particles in new physics beyond the SM, whose masses are unknown.) 
We assume two body decay at each step, $A_i \to b_i B_i$ and $B_i \to c_i C_i$, where $b_i$ and $c_i$ are two visible particles and $C_i$ is the invisible particle. We further assume that $c_1$ and $c_2$ (two charged leptons) are distinguishable, while $b_1$ and $b_2$ (two $b$-tagged jets) are not.
Therefore there is a two-fold ambiguity in pairing $b_i$ and $c_i$. 
Fig. \ref{fig:diagram} shows three possible subsystems in the blue dotted $(b)$, the green dot-dashed $(c)$, and the black solid boxes $(bc)$, respectively.

For the numerical studies in this article, we generate a partonic $t \bar{t}$ dilepton sample, using the \amc~at the 14 TeV LHC with the default parton distribution functions \cite{Alwall:2011uj}. All event samples are generated at leading order accuracy in QCD. The masses of the top quark and the $W$-boson are set to 173 GeV and 80.419 GeV, respectively. 
We take into account the proper finite widths, which often make the top quarks and the $W$-bosons significantly off-shell. In order to reduce the background, we apply the same basic cuts as those used in Ref.~\cite{Choi:2011ys,Debnath:2017ktz}. 
We did not include ISR/FSR in our study but they should be relatively harmless for our problem given the high $b$-tagging efficiency and small fake rates expected at the HL-LHC \cite{CERN-LHCC-2017-021}. 
This setup also allows us to make a fair comparison of our findings against existing results, as most studies in the literature did not consider ISR/FSR.

To simulate detector effects, we follow the  parameterization used in ATLAS detector performances report for the HL-LHC~\cite{ATL-PHYS-PUB-2013-004}. The energy resolution is parameterized by three terms; noise ($N$), stochastic ($S$), and constant ($C$) terms
\begin{align}
\frac{\sigma}{E} =  \sqrt{
\bigg( \frac{N}{E} \bigg)^2 +\bigg( \frac{S}{\sqrt{E}}\bigg)^2  +C^2~,
}
\end{align}
where in our analysis we use $N=5.3$, $S=0.74$ and $C=0.05$ for jets, and $N=0.3$, $S=0.1$, and $C=0.01$ for electrons \cite{Aad:2014nim}, and the energy $E$ is in GeV.   
The muon energy resolution is given by the Inner Detector (ID) and Muon Spectrometer (MS) resolution functions
\begin{align}
\sigma = \frac{ \sigma_{\text{ID}}~ \sigma_{\text{MS} }  } { \sqrt{ \sigma^2_{\text{ID}} + \sigma^2_{\text{MS}   } } }~,
\end{align}
where $\sigma_{\text{ID}}$ and $\sigma_{\text{MS}}$ are defined as 
\bea
\label{eq:MuonSmear}
\sigma_{\text{ID}} &=& E~\sqrt{ \alpha^2_1 + ( \alpha_2 ~ E )^2   } \, ,\\
\sigma_{\text{MS}} &=& E~\sqrt{ \bigg( \frac{\beta_0}{E} \bigg)^2 + \beta^2_1 + (\beta_2 E)^2   }~\;.
\eea
We use $\alpha_1 = 0.023035$, $\alpha_2 = 0.000347$, $\beta_0 = 0.12$, $\beta_1 = 0.03278$ and $\beta_2 = 0.00014$ in our study.
We prepare one million parton-level events and the corresponding one million smeared events for the $b\bar b \ell^+\ell^- + \mptvec$ final state. We denote those smeared events by ``detector-level events'' in the rest of this study.

\subsection{Setup of ML methods}
\label{sec:MLdata}

From the $1$ million events we prepared, we take random selection of $900$k for the training/validation of all of the ML methods (with a 90/10 split), and the remaining 100k events for testing. 

 For all the ML methods in our study, unless otherwise noted, the dimension of the input data is 18, including four momentum of two $b$-tagged jets, two charged leptons and the missing transverse momentum. We order two $b$-tagged jets by their $p_T$ and label them as $b_1$ and $b_2$, and the corresponding correct lepton pairing as $\ell_1$ and $\ell_2$, respectively. Note that $\ell_1$ and $\ell_2$ are not necessarily ordered by their $p_T$. They are the lepton pairing corresponding to the two $p_T$ ordered $b$-jets, $b_1$ and $b_2$. 
 
 We prepare two datasets with labels 1 (correct $b$-$\ell$ pairing) and 0 (incorrect pairing), consisting of ($p_{b_1}$, $p_{\ell_1}$, $p_{b_2}$, $p_{\ell_2}$, $\mptvec$) and ($p_{b_1}$, $p_{\ell_2}$, $p_{b_2}$, $p_{\ell_1}$, $\mptvec$) respectively. In other words, the information about the pairing correctness is encoded in the ordering of the 4-vectors that make up each event. Note that each event is counted twice, once in the correct-pairing dataset and once in the incorrect-pairing dataset. 
We have checked that this re-using / double-counting of events does not affect the performance of the ML methods in any way but helps increase the size of the dataset.
 
 The ML methods are then formulated as binary classifiers (with the usual binary cross-entropy loss and sigmoid activation) between the correct and incorrect pairing datasets. Unlike usual binary classification problems, however, here each event actually produces two scores, one for each ordering of the leptons, i.e. $f(p_{b_1}, p_{\ell}, p_{b_2}, p_{\ell'})$ and  $f(p_{b_1}, p_{\ell'}, p_{b_2}, p_{\ell})$ where $f$ represents the sigmoid output of the ML method. 
Note that previously, we used $\ell_1$ and $\ell_2$ to denote a lepton, which should be paired with $b_1$ and $b_2$, respectively. 
 However, in practice, this truth information is not available and we will have to make an arbitrary pairing of a lepton and $b$-tagged jet.
 For this purpose, we used $\ell$ and $\ell^\prime$ to denote two leptons. 
 Therefore, one of the two scores, $f(p_{b_1}, p_{\ell}, p_{b_2}, p_{\ell'})$ or $f(p_{b_1}, p_{\ell'}, p_{b_2}, p_{\ell})$, is close to 1 (correct pairing), and the other will be close to 0 (incorrect pairing).
 To the extent that the ML method is optimal, we expect that these two scores contain the exact same information, i.e. $f(p_{b_1}, p_{\ell}, p_{b_2}, p_{\ell'})$ and $1-f(p_{b_1}, p_{\ell'}, p_{b_2}, p_{\ell})$ are both equal to the probability that $(b_1,\ell)$ is the correct pairing \cite{neyman1933ix}. Of course, in practice, owing to the finite training data and ML model capacity, these two scores will only be approximate estimates of the true probability. Therefore, going forward we define an averaged score in order to incorporate information from both estimates:
\begin{equation}
P_{{b_1}, {\ell}} = {1+f(p_{b_1}, p_{\ell}, p_{b_2}, p_{\ell'})-f(p_{b_1}, p_{\ell'}, p_{b_2}, p_{\ell})\over 2} \, .
\end{equation}
Note that defined this way, $P$ does have the correct behavior as a binary probability, $P_{b_1,\ell}+P_{b_1,\ell'}=1$.  

To properly compare different methods, we define a common set of metrics as follows. It is natural to take the prediction of the ML method to be the pairing $(b_1,\ell)$ or $(b_1,\ell')$ that returns the higher value of $P$, i.e.\ the ML predicts $(b_1,\ell)$ when $P_{b_1,\ell}>0.5$. 
The fraction of correct predictions provides a baseline measure of the method's performance: the purity at 100\% efficiency.
  
The purity can be improved at the expense of efficiency, by considering a subset of data which passes certain criteria or selection cuts.  For the ML methods, we can accomplish this by requiring the method to be more confident in its prediction, i.e.\ only keeping events for which $P_{{b_1}, {\ell}}>P_c$ or $P_{{b_1}, {\ell'}}>P_c$. Given such a selection, we define  the efficiency ($\epsilon$) and the purity ($P$) to be: \cite{Chen:2008ex,Rajaraman:2010hy,Baringer:2011nh}:
\begin{eqnarray}
\epsilon 	&=& \frac{N_{\rm cuts}}{N_{\rm total}}  ,\label{eq:eff}  \\
&=& \frac{\rm the ~number ~of~ events ~which~ pass ~the ~selection}{\rm the~ total~ number~ of~ events} \,\nonumber \\ 
P  		&=& \frac{N_{\rm correct}}{N_{\rm cuts}} \label{eq:purity}\\
&=& \frac{\rm the ~number ~of~ the~correctly ~identified~events}{\rm the ~number ~of ~events ~that ~passed ~the ~selection}\, \nonumber . 
\end{eqnarray}
Note that in the machine learning literature, the purity is often referred to as precision = ${\rm \frac{TP}{TP + FP}}$, where TP is the number of true positives and FP is the number of false positives. In section \ref{sec:ML} we will produce purity vs.\ efficiency curves (analogous to ROC curves) by varying this threshold $P_c$.

\subsection{Existing methods}

If the individual momentum  of the missing neutrinos (or dark matter candidates) can not be uniquely determined, the next best approach would be to consider some sort of approximation \cite{Kim:2017awi}. For instance, a matrix element method (MEM) can be used to find the most likely values of the invisible momenta
(to be discussed in section \ref{sec:mem}). However, the method itself suffers from the combinatorial problem and is very model-dependent as it requires to fully specify the underlying physics in the consideration (masses, spins, couplings, etc). 

An alternative approach would be to rely only on kinematics such as masses and event topology (without spin or coupling information) and to obtain the invisible momenta by optimizing a suitable kinematic function. But what constitutes a good target function for such an optimization? Several algorithms are proposed depending on what kind of target function is considered: 
kinematic endpoints (section \ref{sec:endpoint}), hemisphere method / recursive jigsaw (section \ref{sec:hemisphere}), topness (section \ref{sec:topness}), kinematic likelihood (KL) fitter (section \ref{sec:KLfitter}), matrix element methods (section \ref{sec:mem}), and analytic reconstruction (section \ref{sec:analytic}). 
In general, methods which invoke fewer assumptions are more robust and model independent, but lead to rather vague conclusions with poor results, while methods with more assumptions give better results, but the methods themselves are fragile as they are not typically applicable to more general cases. This is one of the main reasons why we want to develop as many methods as possible. We also understand kinematics better by comparing how each method works.

In some of these existing kinematic methods, there are natural ways to select a subset of events with higher purity, analogous to the cut $P_{b_1,\ell}>P_c$ or  $P_{b_1,\ell'}>P_c$ described in the previous subsection for the ML methods. For example, in the case of endpoint methods, it is known that a cut on $M_{\rm eff}$ or $H_T$ reduces the number of samples but can improve the accuracy of the pairing prediction \cite{Choi:2011ys}.

\section{Machine Learning approaches}
\label{sec:ML}

 The data collected by high energy physics experiments such as LHC is very complex and very high dimensional. Collider physics analyses could be considered as a series of dimensional reduction processes at several stages. The last stage is likely to involve the reconstructed objects. Even then, the dimension of the input data is still quite large -- proportional to the number of reconstructed particles. It is a difficult task to understand the full correlations of the high dimensional data; this has motivated the consideration of suitable kinematic variables that capture the salient features of the initial data, and it also motivates the study of ML-based methods where this kind of feature engineering is automated directly from the initial data.

 In the following subsections, we will explore
 various algorithms such as tree-based methods (section \ref{sec:tree}), deep neural networks (section \ref{sec:dnn}), recurrent neural networks (section \ref{sec:rnn}), attention-based networks (section \ref{sec:attention}) and Lorentz boost network (section \ref{sec:LBN}). We discuss each method very briefly without going into details. We refer to Refs. \cite{hastie01statisticallearning,bishop:2006:PRML,Feickert:2021ajf,Guest:2018yhq} for more details of various machine learning algorithms and Refs. \cite{Guest:2018yhq,Feickert:2021ajf,Karagiorgi:2021ngt} for machine learning in high energy physics. 
 All codes used in this paper are publicly available at \cite{codes}.

\subsection{Tree-based methods}
\label{sec:tree}
\begin{figure*}[ht!]
\begin{center}
\includegraphics[width=0.43\textwidth]{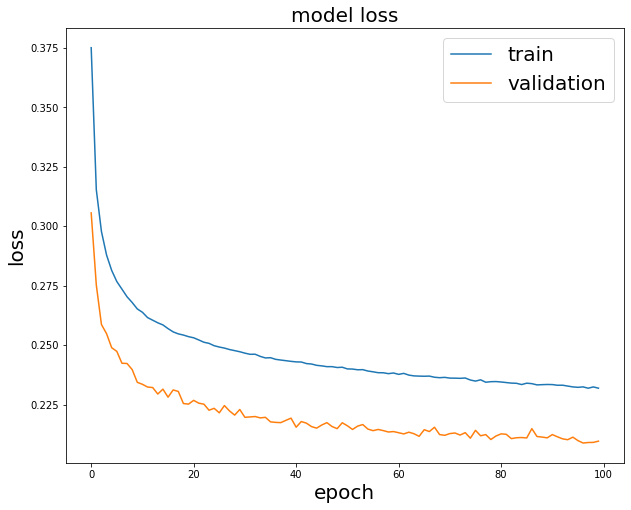} \hspace*{0.5cm}
\includegraphics[width=0.43\textwidth]{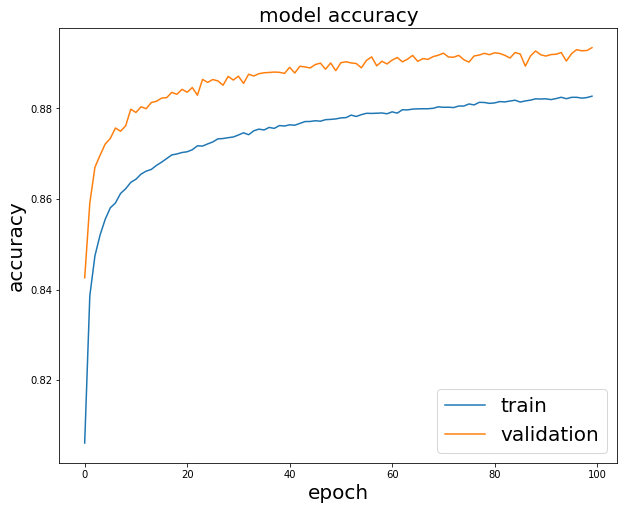} \\
\includegraphics[width=0.43\textwidth]{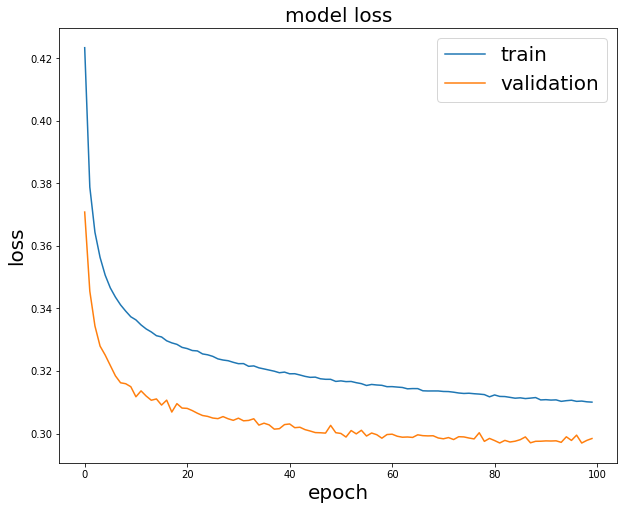}\hspace*{0.5cm}
\includegraphics[width=0.43\textwidth]{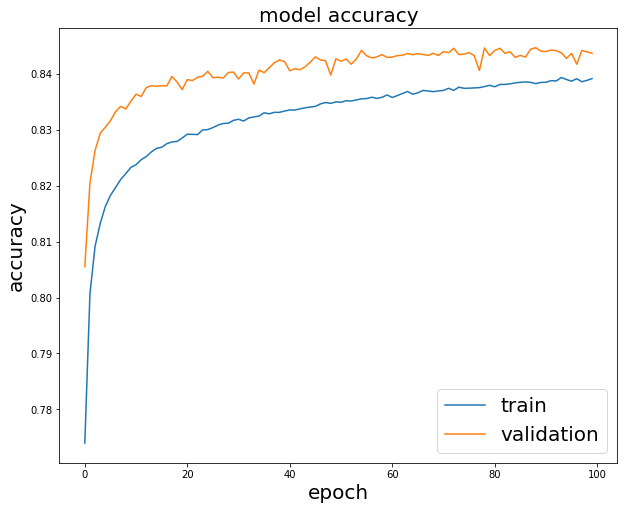}
\caption{DNN loss (left) and accuracy (right) for the parton-level (top) and the detector-level (bottom) events.}
\label{fig:dnnloss}
\end{center}  
\end{figure*}

As a rudimentary ML baseline, we consider two common tree-based methods: random forest and boosted decision trees. These are shallow ML methods that are popular in HEP and can handle a modest number of inputs. For more about these methods, we refer the reader to \cite{Hocker:2007ht,scikit-learn}.

First we tried the tree-based methods with four momentum information only (18 dimensional input).
In our analysis, the forest includes $2000$ trees, with each leaf having at most $0.01\%$ of the entire training data. The fraction of training set used to train each tree does not have a significant impact in our result. The random forest algorithm is implemented using the Scikit-learn python library \cite{scikit-learn}. 
By taking the combination which returns the higher score as the correct combination, we find that the random forest classifies $83.4\% $ of the data correctly for the parton-level events, and $81.2\%$ for the detector-level events, respectively (for 100\% efficiency).

In general, boosted decision trees (BDT) outperform most other tree based methods. 
In our analysis, the implementation of the algorithm is done using the XGBoost library in python \cite{Chen:2016:XST:2939672.2939785}. 
We find that the BDT classifies $86.1\% $ of the data correctly for the parton-level events and $82.4\% $ for the detector-level events, respectively. 
This is comparable to or better than the existing methods discussed in Appendix \ref{sec:existingmethods}. A good performance with the detector-level data is especially notable. 

We repeated the same analysis, with boosted trees and random forest, including the kinematic variables ($m_{b\ell}$, $M_{2Ct}$, $M_{2CW}$, $\Delta R_{b\ell}$, topness) defined in Appendix and obtained a higher purity 90.4\% and 90.1\% for parton-level events and 84.1\% and 84\% for detector-level events, respectively.  The fact that including kinematic variables improved the performance of these algorithms indicates that the random forest and BDT did not catch important kinematic features during the training -- possibly a sign that shallow ML methods are insufficiently expressive to fully automate feature learning.

Finally, we also tried training random forests and BDTs on kinematic variables only (without four momenta). The results do not change significantly; we obtained 89.5\% and 83.4\% purity using boosted trees and 89.7\% and 83.7\% purity using random forest, respectively. Evidently the four-vectors are not adding much to the performance beyond the kinematic variables, for these shallow ML methods.

\subsection{Deep neural networks}
\label{sec:dnn}
\begin{figure*}[t]
\begin{center}
\includegraphics[width=0.43\textwidth]{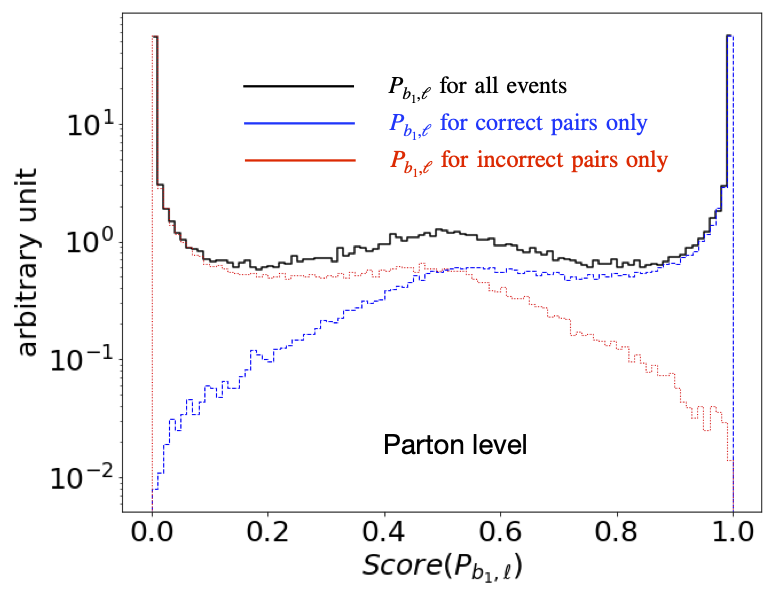}\hspace*{0.5cm}
\includegraphics[width=0.43\textwidth]{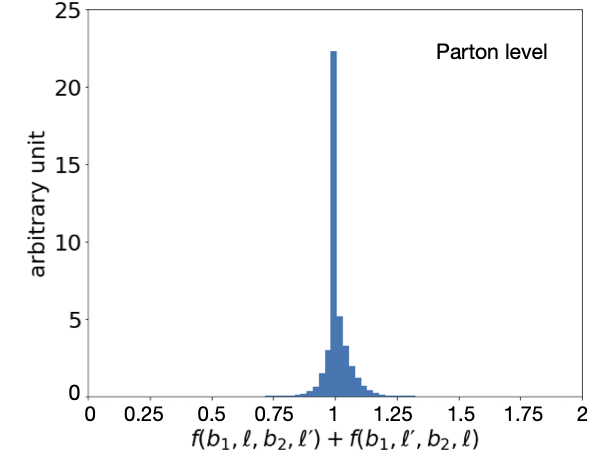} \\
\includegraphics[width=0.43\textwidth]{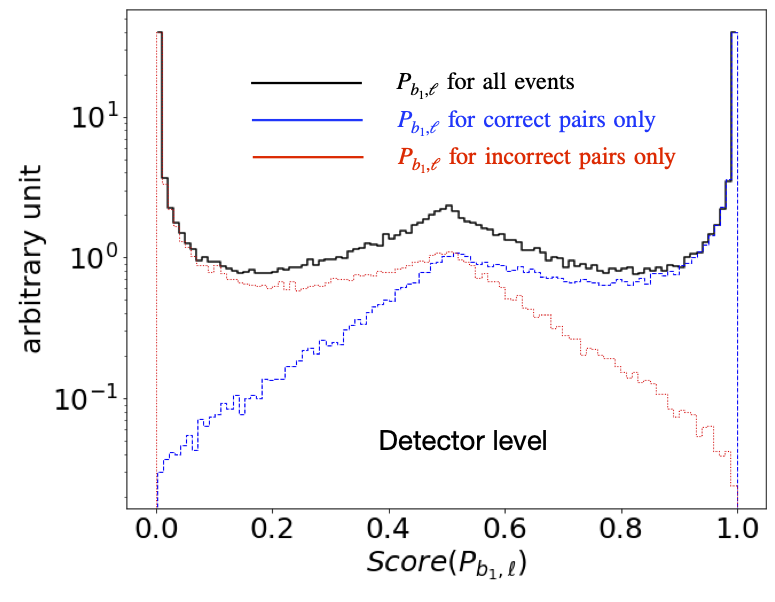}\hspace*{0.5cm}
\includegraphics[width=0.43\textwidth]{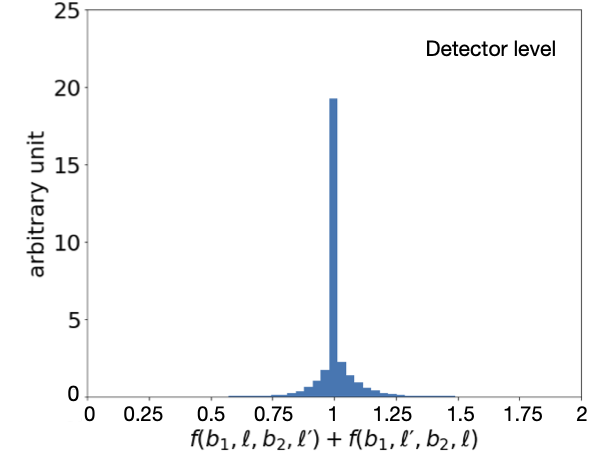}
\caption{
Distributions of the DNN scores (left), and the sum of two scores ($f(p_{b_1}, p_{\ell}, p_{b_2}, p_{\ell'}) + f(p_{b_1}, p_{\ell'}, p_{b_2}, p_{\ell})$) (right) for parton-level events (top) and detector-level events (bottom). 
The score $P_{b_1,\ell}$ for all events is shown in black, while $P_{b_1,\ell}$ for which $b_1,\ell$ is the correct (incorrect) pairing is in blue (red) and peaked close to 1 (0).}
\label{fig:score}
\end{center}  
\end{figure*}

Next we consider a fully-connected deep neural network (DNN).
The input layer is followed by $3$ hidden layers of $512$ neurons each, with ReLU activation functions in between. The last hidden layer is connected to a single output neuron with the sigmoid activation function, to match the target binary label of $0$ (for the incorrect pair) or $1$ (for the correct pair). 

To prevent overfitting during training, we consider a $25\%$ dropout after each hidden (DNN, LSTM and LBN) layer. We find this dropout value makes the training and validation accuracy converge relatively well as shown in Fig. \ref{fig:dnnloss}. (As is generally the case, dropout is only used during training and not during validation/testing, which causes the training accuracy/loss to appear worse than the validation accuracy/loss over the training history.) Batch Normalization is applied before each dropout. We used the Adam optimizer with a learning rate of $10^{-3}$ to minimize the binary cross entropy loss function for all NNs that considered in this paper.
The whole network structure is implemented using the Keras library \cite{chollet2015keras} and the best model is selected based on the validation loss. 

Fig.~\ref{fig:score} shows the score distributions ($P_{b_1 , \ell}$ in black for all events, $P_{b_1 , \ell}$ in blue for which $(b_1,\ell)$ is the correct pairing and $P_{b_1 , \ell}$ in red for which $(b_1,\ell)$ is the incorrect pairing) in the left and the sum of two raw DNN outputs ($f(p_{b_1}, p_{\ell}, p_{b_2}, p_{\ell'}) + f(p_{b_1}, p_{\ell'}, p_{b_2}, p_{\ell}) $) in the right panel for parton-level (top) and detector-level events (bottom). 
As expected, the sum of the two DNN outputs is approximately (but not exactly) equal to 1, and the 
$P_{b_1 , \ell}$ score for which $(b_1,\ell)$ is the correct (incorrect) pairing peaks near 1 (0). 
We find the DNN predicts correctly $89.5\%$ of the time at parton-level and $84.5\%$ at the detector-level, which are are very similar to the accuracy (right panel) in Fig. \ref{fig:dnnloss}. 

As a cross-check, we have tried a different approach. 
Instead of taking both correct and incorrect combinations (i.e., instead of using the same event twice) and producing two scores $P_{b_1,\ell}$ and $P_{b_1,\ell^\prime}$ for each event, we have tried to produce a single score by $p_T$ ordering the leptons in event, and labeling each event according to whether this $p_T$ ordering produced the correct pairing or not. With the single score, we find a minor difference for the parton-level purity (88\% for 100\% efficiency), while the detector-level purity (84\%) remains very similar to the case with the two scores ($P_{b_1,\ell}$ and $P_{b_1,\ell^\prime}$). Since our first method gives a slightly better outcome, we use the two-score scheme for the remaining NNs.

%
\begin{figure*}[bt!]
\begin{center}
\includegraphics[width=0.9\textwidth]{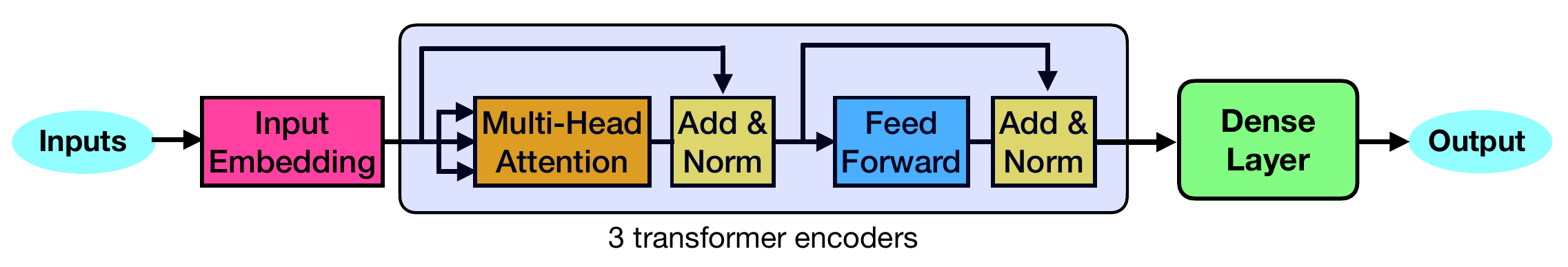}
\caption{Schematic diagram of the attention network in our analysis.\label{fig:attention}} 
\end{center}  
\end{figure*}

We also find that the result remains approximately the same when we exclude $\mptvec$ information in the input, keeping the $4$-momentum of the visible particles only. This is to be expected since the $\mptvec$ in our simulation is simply negative of the sum of the transverse momentum of all visible particles, implying that the $\mptvec$ does not add new information. We also notice that, unlike the tree-based methods, including more inputs such as the kinematic variables or topness value does not change the result of neural networks significantly.
This observation indicates that NN learns the (high-level) kinematic features efficiently from the low-level features (four momenta), unlike tree-based methods.
NN with kinematic variables only (without four momenta) leads to slightly lower but comparable purity 89.75\% and 83.98\% for the parton-level and the detector-level events, respectively. 
Finally, one can improve the purity at the cost of statistics.
For example, we can cut on the value of $P_{{b_1}, {\ell}}$ (keeping the events for which $P_{{b_1}, {\ell}}>P_c$ or $P_{{b_1}, {\ell'}}>P_c$) targeting 99\% purity, which leads to the efficiency of 76.2\% and 54.3\% for parton-level and detector-level events, respectively. More details will be discussed in section \ref{sec:comparison}.

\subsection{Long Short-Term Memory}
\label{sec:rnn}

Next we consider whether the NN can extract more information from the data by treating it as a sequence. We focus on the Long Short-Term Memory (LSTM) architecture \cite{Hochreiter:1997yld},
which was originally developed to overcome the vanishing gradient problem that often arises when training more traditional recurrent neural networks (RNNs). In collider physics, RNNs have been applied to study jet tagging problems, where the input is a sequence of jet kinematic information \cite{Guest:2018yhq,deLima:2021fwm}. In dileptonic $t\bar t$ production, one can consider the time-ordering of the final state particles. Two $b$-quarks would appear before two leptons or two neutrinos. Although the decay is somewhat short, it is worth investigating the performance of the LSTM and comparing against that of the DNN.

We implement the LSTM using {\tt TensorFlow} \cite{abadi2016tensorflow}. The 16 dimensional input vector (without the missing transverse momentum) made up of four momenta of four particles (four sequences) are fed into two LSTM layers of 256 dimensions of the output space, followed by one DNN layer with 128 neurons. We use the same optimizer, loss function and dropout used for DNN in section \ref{sec:dnn}. The number of trainable weights in the LSTM network is 1 mil, which is comparable to the size of the DNN considered in the previous subsection. We obtain the purity of 89.27\% and 83.95\% for the parton-level and detector-level events, which are similar to those with DNN.

\subsection{Attention-based network} 
\label{sec:attention}

Neural networks with attention is a technique that imitates cognitive attention. Attention is the ability to choose and concentrate on relevant stimuli, and respond accordingly \cite{NIPS2017_3f5ee243}. 
The effect of attention module enhances the important parts of the input data and fades out the less important parts such that the network devotes more computing power on more relevant part of the data. Which part of the data is more important than the others depends on the problem and is learned via training. Attention-based networks are used in a wide variety of machine learning models, including in natural language processing and computer vision.

Recently, the combinatorial problem in $t\bar t$ production with fully-hadronic top decays has been examined in detail using attention-based neural networks \cite{Shmakov:2021qdz,Fenton:2020woz,Lee:2020qil}. The authors showed that the performance significantly improved over the traditional kinematic methods. 
In this section, we apply an attention-based network to the two-fold ambiguity in the dilepton production. 
Our NN is based on the standard self-attention network, which is relatively simpler than the architecture used in Ref. \cite{Shmakov:2021qdz,Fenton:2020woz,Lee:2020qil}. There are two reasons for this: both our input and output are simpler.  For input, we only need to consider the two $b$-tagged jets without worrying about all the extra jets, which makes our input fixed length, compared to the variable lengths in the fully hadronic final state. As for the output, the fully hadronic channel is more complicated because one needs to identify which jets are selected as well as which jets form the top quark / $W$ boson, while in our case whatever label we obtain is essentially a two-fold ambiguity, and it is mathematically equivalent to using a binary label.

We implemented the self-attention mechanism in our study by using the transformer encoder layer from Pytorch \cite{NEURIPS2019_9015}. The input to this network is similar to that of the LSTM network, which are the 4-momentum of the $4$ visible particles. Each particle momentum is first fed into dense embedding layers of dimensions $8$, $32$ and $64$ respectively. The embedded vectors are fed into three transformer encoder layers, where the inputs first flow through a multi-head self-attention layer with $4$ heads, and then into a feed forward layer, with residual connection over each of the layers . The $4$ output vectors of those layers with dimension $64$ are then flattened, before feeding into another set of dense layers of dimension $64$ and $1$, with the last layer being the output layer of the network that has a sigmoid activation. 
We obtain 89.8\% purity for parton-level and 84.4\% for detector-level events, respectively, which are comparable to results using DNN and LSTM.

\begin{figure*}[t!]
\begin{center}
\includegraphics[width=0.9\textwidth]{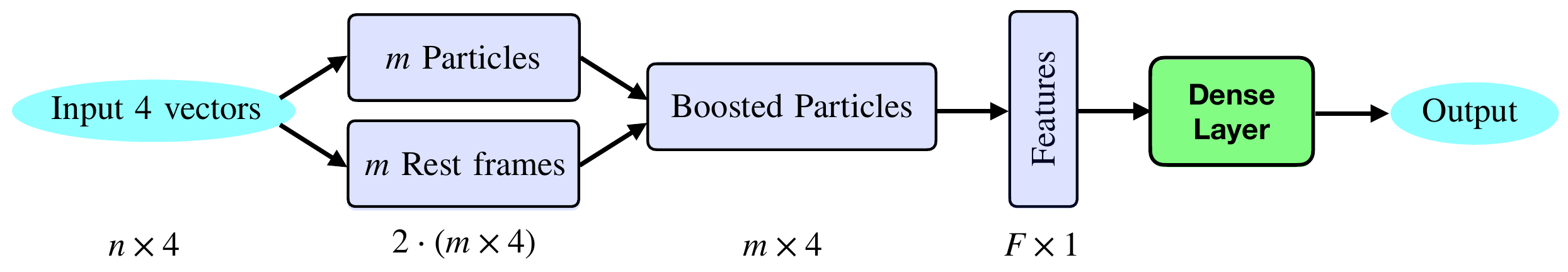}
\caption{Schematic diagram of the Lorentz Boost Network.\label{fig:lbn} See Refs. \cite{Erdmann:2018shi,LBNtth} for more details.} 
\end{center}  
\end{figure*}
%

\subsection{Lorentz Boost Networks}
\label{sec:LBN}

Lorentz Boost Networks (LBNs) are motivated by particle kinematics in the rest frames of various particle combinations. 
The LBNs take four-momentum of $n$ final state particles as an input and create $m$ arbitrary combinations of particles, which are boosted to the rest frame of $m$ different combinations of particles. The set of $m$ particle combinations and the set of $m$ rest frames are constructed by a linear combination of input momenta determined by two trainable $n$ by $m$ matrices. This step creates four-momenta of $m$ particles (boosted to the rest frames), which are used to calculate the high level variables such as invariant mass or the angle between any particle pairs.
The parameter $m$ is taken to be a hyperparameter of the network. These high level variables are combined with the four momentum of the boosted particles and fed into a simple DNN classification network. The LBNs have been applied for the semi-leptonic channel of $t\bar t h$ production \cite{Erdmann:2018shi,LBNtth} and dilepton $t\bar t$ production \cite{LBN}.
 
As the LBN architecture takes advantage of particle kinematics in various rest frames of composite particles, it is worth investigating the two-fold ambiguity using the LBN. 
Since the complete kinematic information of all the final state particles is unknown in our study, we use the momentum of the $4$ visible particles (two b-tagged jets and two leptons) and the missing transverse momentum with zeros in the $E$ and $p_z$ component. Note that LBN does not use the mass information explicitly.  

We use the TensorFlow implementation of the Lorentz Boost Network, which is available from {\verb|https://git.rwth-aachen.de/3pia/lbn|}. 
The schematic diagram of the Lorentz Boost Network is shown in Fig. \ref{fig:lbn}. 
We feed the 20 dimensional input momenta into the LBN layer. We set the number of particles ($n$) and rest-frame combinations ($m$) that LBN builds to $n=m=8$, which leads to the optimal result.
The LBN layer returns the $F=69$ dimensional output made up of 6 kinematic features ($E$, $p_T$, $\eta$, $\phi$, mass, $\cos\theta_{ij}$, which is the decay angle of $i$-th particle in the $j$-th rest frame) for those 8 particles, which are fed into two DNN layers with 512 neurons for each. 

At parton-level with $4$-momentum input to the classifier, we obtain an accuracy of $89.8\%$. With the detector-level events, we obtain an accuracy $86\%$. For other values of $m$, we obtained slightly worse result. For example, for $m=4$, we obtained 88\% purity at parton-level and 85\% at detector-level, respectively.

\subsection{Comparison of different methods}
\label{sec:comparison}

Now we compare the performance of each method in finding the correct and incorrect combination of a $b$-quark and a lepton for the dilepton $t\bar t$ production. Table \ref{table:summary} summarizes the purity ($P$) and the efficiency ($\epsilon$) for various ML approaches discussed in section \ref{sec:ML} and the existing methods described in Appendix \ref{sec:existingmethods}. All methods use four momenta of four visible particles and the missing transverse momentum as basic inputs. 
Some examples use kinematic variables \big ($m_{b\ell}$, $M_{2CW}^{(b\ell)}$,  $M_{2Ct}^{(\ell)}$, topness, $\Delta R_{b\ell}$\big ) in addition to the basic inputs, except for the cases ``with kinematic variables only'', where four momenta are omitted in the inputs.
\begin{table*}[t]
\def\arraystretch{1.5}
\begin{center}
\scalebox{1.08}{
\begin{tabular}{||c||c ||c | c|| c| c||} 
 \hline
 \multirow{2}{*}{Algorithm} 
  &~  \multirow{2}{*}{Section}~ & \multicolumn{2}{c||}{~parton-level~ } & \multicolumn{2}{c||}{~detector-level~ }\\ \cline{3-6} 
   & & P & $\epsilon$  & P & $\epsilon$\\ \hline \hline 
Endpoints method I \big ($m_{b\ell}$, $M_{2CC}^{(b\ell)}$,  $M_{2CC}^{(\ell)}$\big )& \ref{sec:endpoint} & 0.816 & 1 & 0.789 & 1 \\ \hline
Endpoints method II ({\color{cyan}\ding{72}})  \big ($m_{b\ell}$, $M_{2CW}^{(b\ell)}$,  $M_{2Ct}^{(\ell)}$\big )& \ref{sec:endpoint} & ~\,0.957\,~ & \,\,0.769\,\, & ~\,0.874\,~ & \,0.742\,\\ \hline
Hemisphere method&\ref{sec:hemisphere}& 0.78 & 1 & 0.77 & 1\\ \hline
Recursive Jigsaw &\ref{sec:hemisphere}& 0.762   & 1 &  0.757 & 1\\ \hline
Topness method  ({\color{asparagus}\ding{117}}) &\ref{sec:topness}& 0.869 & 1 & 0.814 & 1\\ \hline
KLfitter ({\color{magenta}\ding{115}})&\ref{sec:KLfitter}& 0.866 & 1 & 0.776   & 1 \\ \hline 
Matrix element method &\ref{sec:mem}& 0.847   & 1 & 0.817 & 1\\ 
\hline
\hline
Boosted decision tree &\ref{sec:tree}& 0.861 & 1 & 0.824 & 1\\ \hline
BDT with kinematic variables&\ref{sec:tree}& 0.904 & 1 & 0.841 & 1\\ \hline
BDT with kinematic variables only&\ref{sec:tree}& 0.895  & 1 & 0.834  & 1\\ \hline
Random Forest &\ref{sec:tree}& 0.834 & 1 & 0.812 & 1\\ \hline
Random Forest with kinematic variables&\ref{sec:tree}& 0.901 & 1 & 0.840 & 1\\ \hline
~~~~~Random Forest with kinematic variables only~~~~~&\ref{sec:tree}& 0.897    & 1 &    0.837 & 1\\ \hline
DNN &\ref{sec:dnn}& 0.895 & 1 & 0.845 & 1\\ \hline
DNN ({\color{red}\ding{53}}) &\ref{sec:dnn}& 0.990 & 0.721 & 0.990 & 0.543\\ \hline
DNN with kinematic variables &\ref{sec:dnn}& {\bf 0.907}  & 1 & 0.846  & 1\\ \hline
DNN with kinematic variables only &\ref{sec:dnn}& 0.898 & 1 & 0.839 & 1\\ \hline
Long Short-Term Memory &\ref{sec:rnn}& 0.893  & 1 & 0.839  & 1\\ \hline
Attention Network &\ref{sec:attention}& 0.898 & 1 & 0.844 & 1\\ \hline
Lorentz Boost Network &\ref{sec:LBN}& 0.898 & 1 & {\bf 0.860} & 1\\ \hline
\end{tabular}
}
\end{center}
\caption{Summary of the purity ($P$) and the efficiency ($\epsilon$) for various methods discussed in section \ref{sec:ML} and Appendix \ref{sec:existingmethods}. Basic inputs are four momenta of two $b$-quarks, two leptons and the missing transverse momentum. 
Some examples use kinematic variables \big ($m_{b\ell}$, $M_{2CW}^{(b\ell)}$,  $M_{2Ct}^{(\ell)}$, topness, $\Delta R_{b\ell}$\big ) in addition to the basic inputs. 
\label{table:summary}
}
\end{table*}

Most methods lead to similar results. 
Especially, if BDT and Random Forest take advantage of additional kinematic features, their performance is comparable to that of DNN. However, in the case of DNN, we have not observed such improvement even with the additional features. We believe that this is due to the flexibility and the efficiency of NN, which learns the high-level features (kinematic variables) from the lower-level data (four momenta inputs) and therefore NN does not need kinematic features as additional inputs. Even with different network structures, the results did not change significantly.

Table \ref{table:summary} shows that, when using some machine learning algorithms such as BDT or DNN, the methods with ``kinematic variables only'' lead to comparable results, surpassing the performance of the traditional use of the kinematic variables (topness, endpoints or hemisphere). This observation tells us that ML algorithms are able to find non-trivial correlation among these kinematic variables efficiently. 
However, the same algorithms do not improve with both four momentum and kinematic variables as inputs. 
We also notice that most ML methods lead to the purity $\sim 90\%$ and $\sim 85\%$, for the parton-level and detector-level events.

\begin{figure*}[t]
\begin{center}
\includegraphics[width=0.46\textwidth]{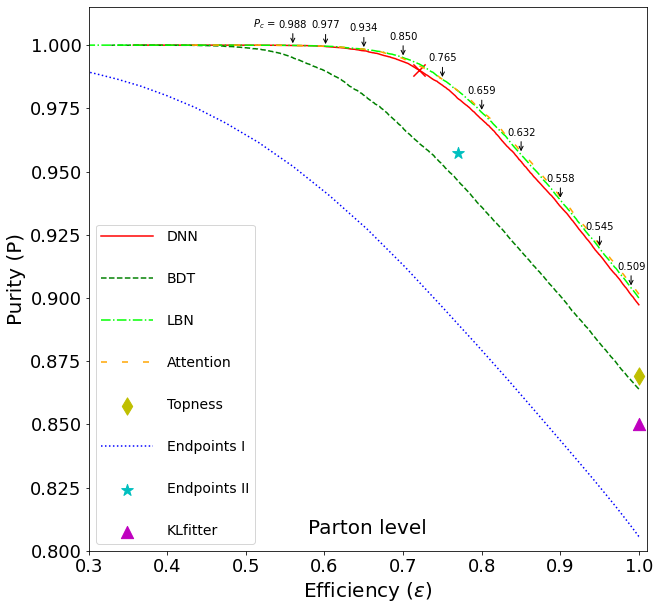} \hspace*{0.4cm}
\includegraphics[width=0.46\textwidth]{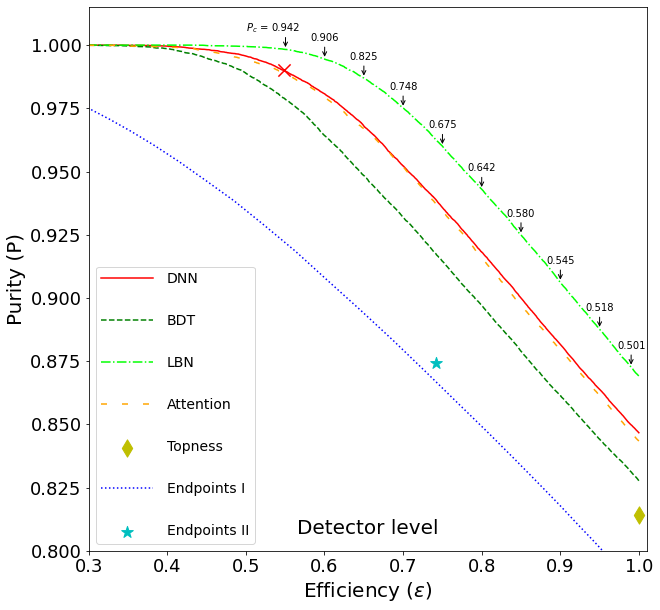}
\caption{Purity ($P$) vs efficiency ($\epsilon$) curves for parton-level events (left) and detector-level events (right). 
As illustration, we show the values of $P_c$ (for $\epsilon \in (0.55, 0.99)$ with 0.05 interval), which are used when cutting on the LBN score $P_{{b_1}, {\ell}}>P_c$ or $P_{{b_1}, {\ell'}}>P_c$.
\label{fig:ROC}}
\end{center}  
\end{figure*}

Fig. \ref{fig:ROC} summarizes the performance of each method in the purity-efficiency plot for the parton-level events (left) and the detector-level events (right), respectively, taking four momenta as inputs (except for the two endpoint methods, which use the kinematic variables as inputs.). Note that the purity is defined as precision, which is the ratio of the number of true positives to the number of events that pass the selection cuts (Eq. \ref{eq:purity}). The efficiency is the ratio of the number of events that pass the selection cuts to the total number of events (Eq. \ref{eq:eff}). The selection criteria that define the various purity-efficiency curves are described in section~\ref{sec:setup}. Results using conventional methods such as the topness, endpoints method II, and KL fitter methods are marked as {\color{asparagus}\ding{117}}, {\color{cyan}\ding{72}} and {\color{magenta}\ding{115}}, respectively.  (The efficiency of endpoint method II is reduced due to the presence of unresolved events.) 
The {\color{red}\ding{53}} mark is shown as a reference to 99\% purity for DNN. For the endpoint method I ($m_{b\ell}$, $M_{2CC}^{(b\ell)}$, $M_{2CC}^{(\ell)}$ without mass information), we cut on the transverse mass ($M_T^{t\bar t} = \sqrt{ \hat{s}}_{min}$) of the entire $t\bar t$ system to make the purity-efficiency curve \cite{Choi:2011ys,Barr:2011xt,Konar:2010ma,Konar:2008ei}. 

DNN (solid-red), attention network (long-dashed orange) and LBN (dot-dashed bright-green) show the best performance with LBN being better for detector-level events. We did not find any difference between attention-based network and DNN.  Results using BDT and endpoint method I are shown in dark green-dashed and blue dotted curve, respectively. 

Taking 99\% (95\%) as a benchmark purity, the efficiencies of ML methods are [0.284, 0.599, 0.721, 0.732, 0.734] ([0.560, 0.758, 0.861, 0.873, 0.867]) for parton-level and [0.173, 0.495, 0.548, 0.541, 0.634] ([0.449, 0.645, 0.707, 0.0705, 0.779]) for detector-level for endpoint I, BDT, DNN, Attention network and LBN, respectively.
We see that deep learning and the LBN method in particular brings impressive efficiency gains at these high levels of purity, which could have major benefits for physics analyses that rely on resolving the combinatorial ambiguity in dileptonic $t\bar t$.

\section{Finding the correct partition without any mass information}
\label{sec:bsm}

We have investigated various methods to resolve the two-fold ambiguity in the dilepton $t\bar t$ production. Some methods use explicit mass information, while others do not. 
In this section, we consider the two-fold ambiguity in the same topology as in Fig. \ref{fig:diagram} but with the unknown mass spectrum. As an illustration, we fix the mass of $A$, $m_A=500$ GeV and  scan over two dimensional mass parameters, ($m_B$, $m_C$) with the mass constraint, $0< m_C < m_B < m_A$. We will use the endpoint method I, hemisphere method and neural networks. For the endpoint method I, we compute \big ($m_{b\ell}$, $M_{2CC}^{(b\ell)}$,  $M_{2CC}^{(\ell)}$\big ) without using explicit masses in the minimization. Here the $M_{2CC}$ is defined as 
\begin{eqnarray}
M_{2CC} &\equiv& \min_{\vec{q}_{1},\vec{q}_{2}}\left\{\max\left[M_{P_1}(\vec{q}_{1},\tilde m),\;M_{P_2} (\vec{q}_{2},\tilde m)\right] \right\}  ,~~~
\label{eq:m2CCdef}\\
 \mptvec  &=&  \vec{q}_{1T}+\vec{q}_{2T} \nonumber \, ,\\
M_{A_1}&=& M_{A_2} \nonumber  \, ,\\
M_{B_1}&=& M_{B_2} \nonumber \, ,
\end{eqnarray}
where $M_{2CC}^{(b\ell)}$ and $M_{2CC}^{(\ell)}$ are $M_{2CC}$ variable applied to $(b\ell)$ and $(\ell)$ subsystem, respectively. They are similar to $M_{2CW}^{(b\ell)}$ and $M_{2Ct}^{(\ell)}$, but only the mass-equality conditions ($M_{A_1}=M_{A_2}$ and $M_{B_1}=M_{B_2}$) are imposed. However no numerical values are used during minimization. Note that these variables satisfy, $M_{T2} = M_2 \leqslant  M_{2CC} \leqslant  M_{2CC}~ {\rm ( with~mass~input)} \leqslant  \max(M_{P_1}, M_{P_2}) =m_A$, where $M_{P_i}$ is the mass of the mother particle in the $i$-th side, and $M_{2}$ is the same as Eq. (\ref{eq:m2CCdef}) without the two mass constraints. These kinematic variables for the correct combination are bounded by their maximum endpoint, while the incorrect combination can vary and could violate the kinematic endpoint. Therefore the mass variables for the correct combination tend to be smaller than values for the incorrect combination. By requiring that the partition which gives  more  ``smaller'' values as the ``correct'' one,  we  can  resolve  two-fold  ambiguity without using mass information. Note that there are no unresolved events, leading to 100\% efficiency. 

\begin{figure*}[t!]
\begin{center}
\includegraphics[width=0.413\textwidth]{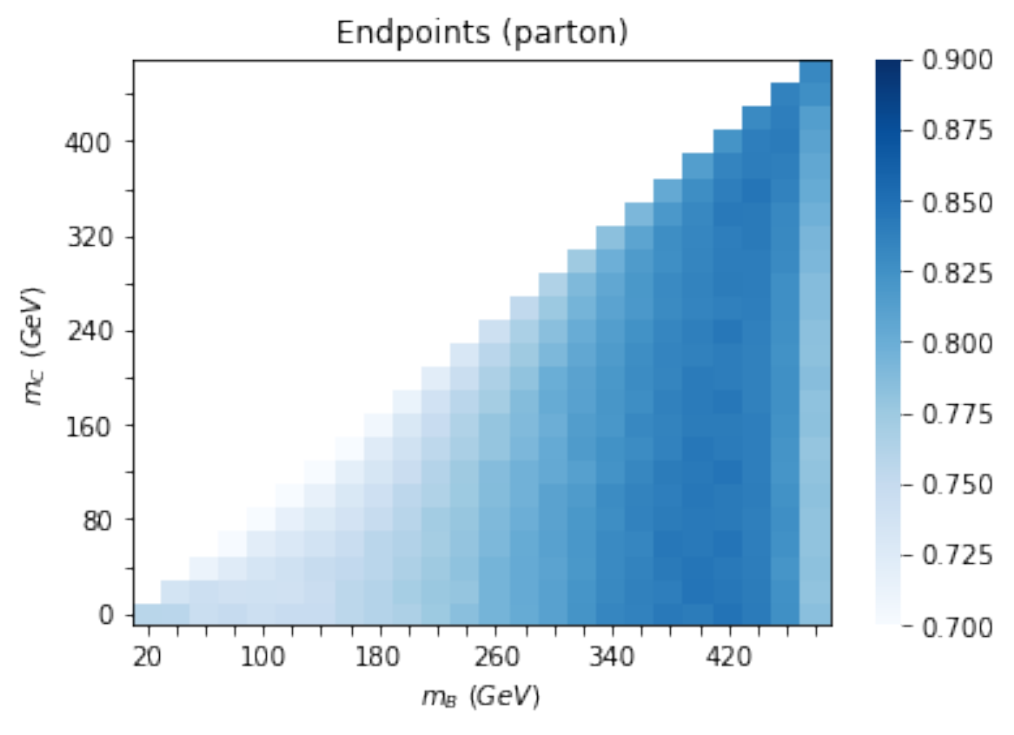} \hspace{0.5cm}
\includegraphics[width=0.413\textwidth]{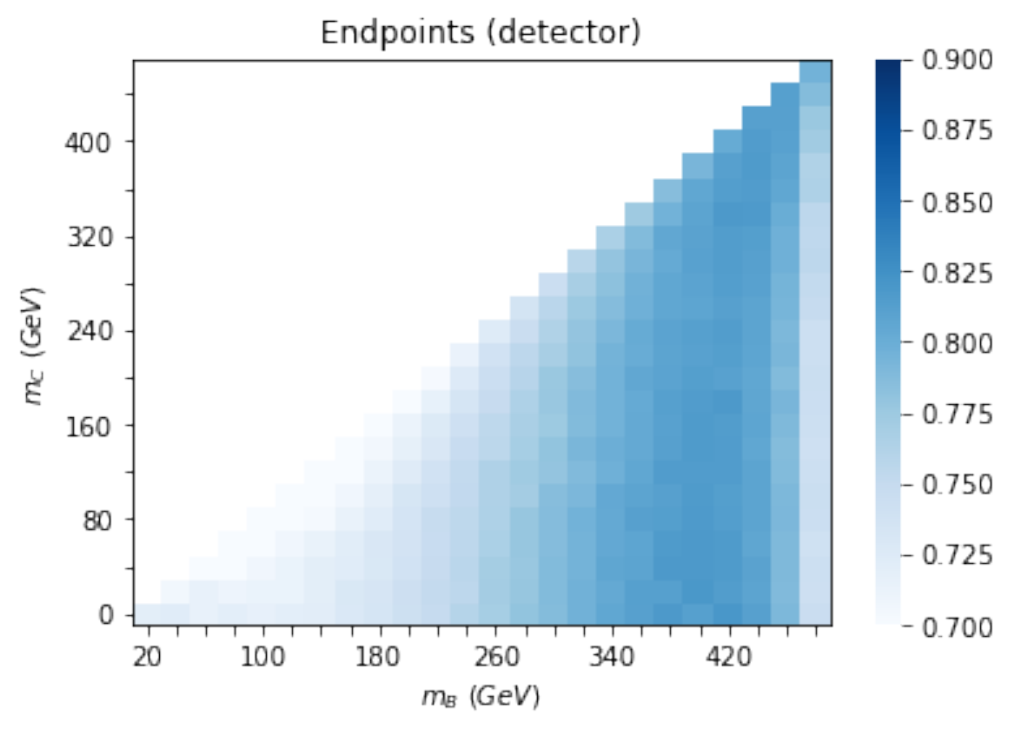} \\
\includegraphics[width=0.413\textwidth]{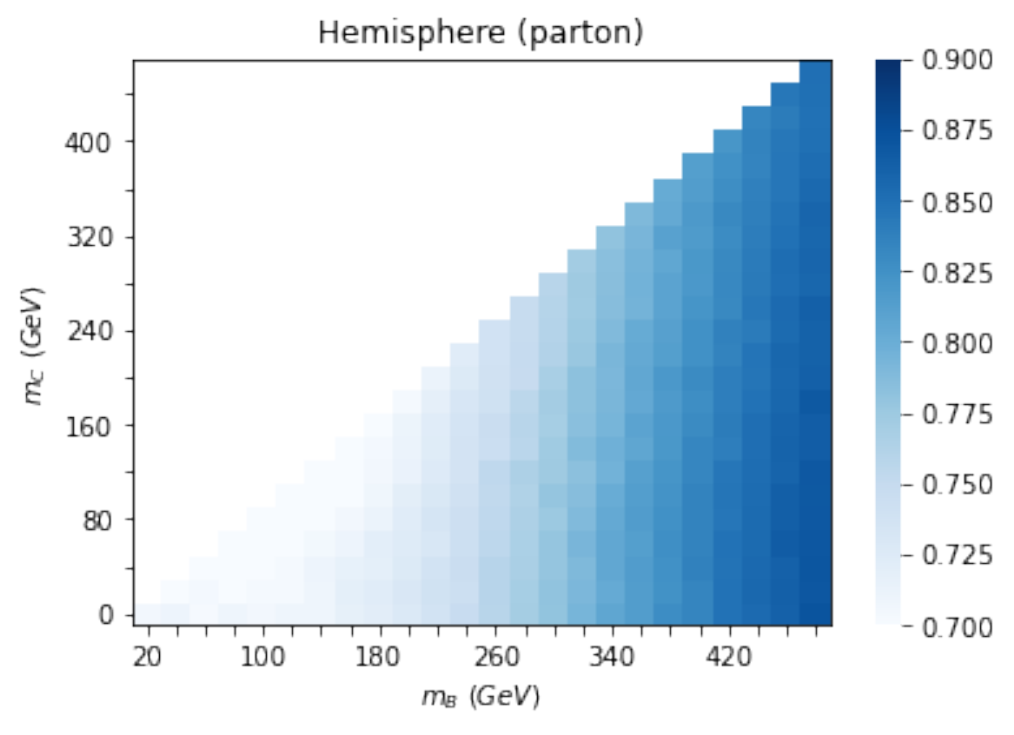} \hspace{0.5cm}
\includegraphics[width=0.413\textwidth]{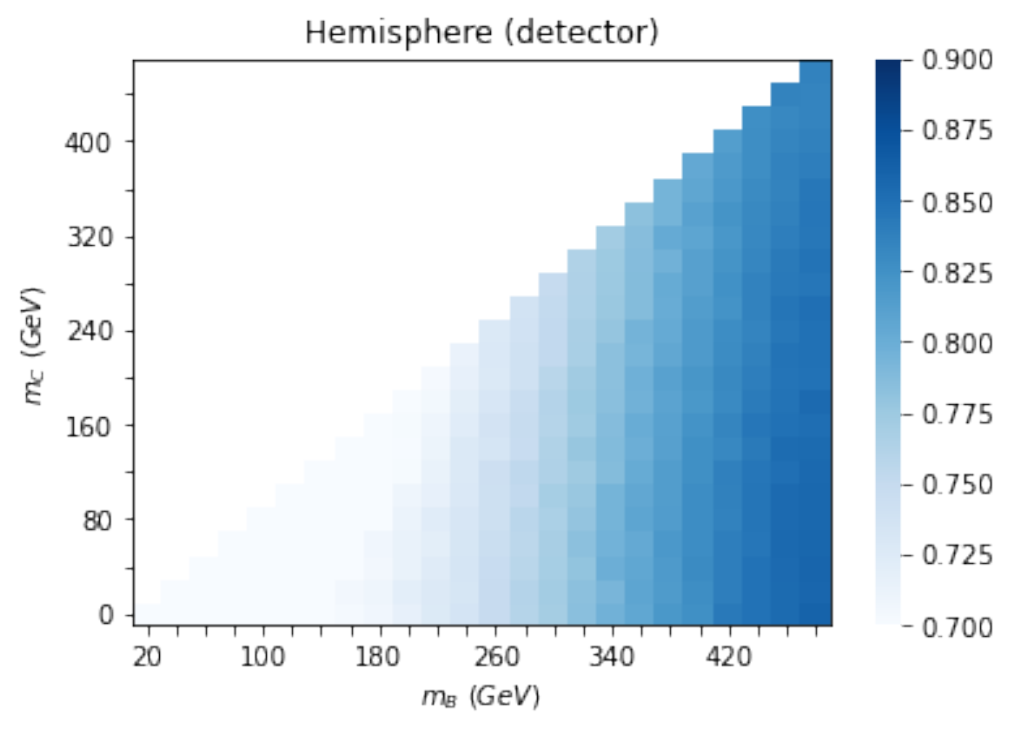} \\
\includegraphics[width=0.413\textwidth]{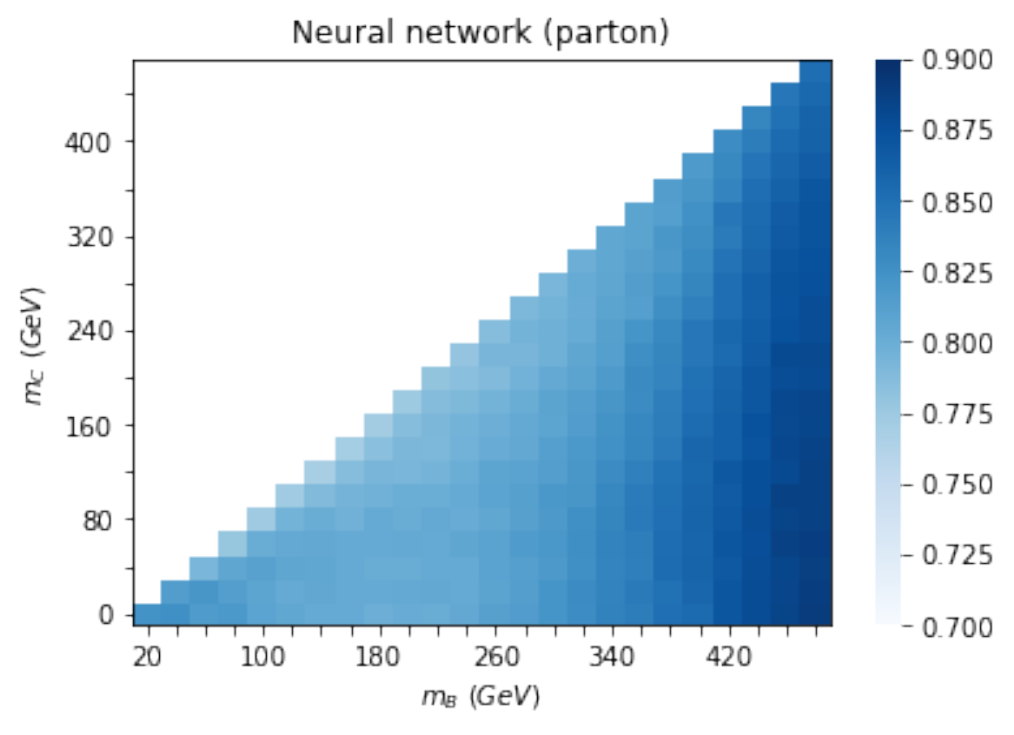}\hspace{0.5cm}
\includegraphics[width=0.413\textwidth]{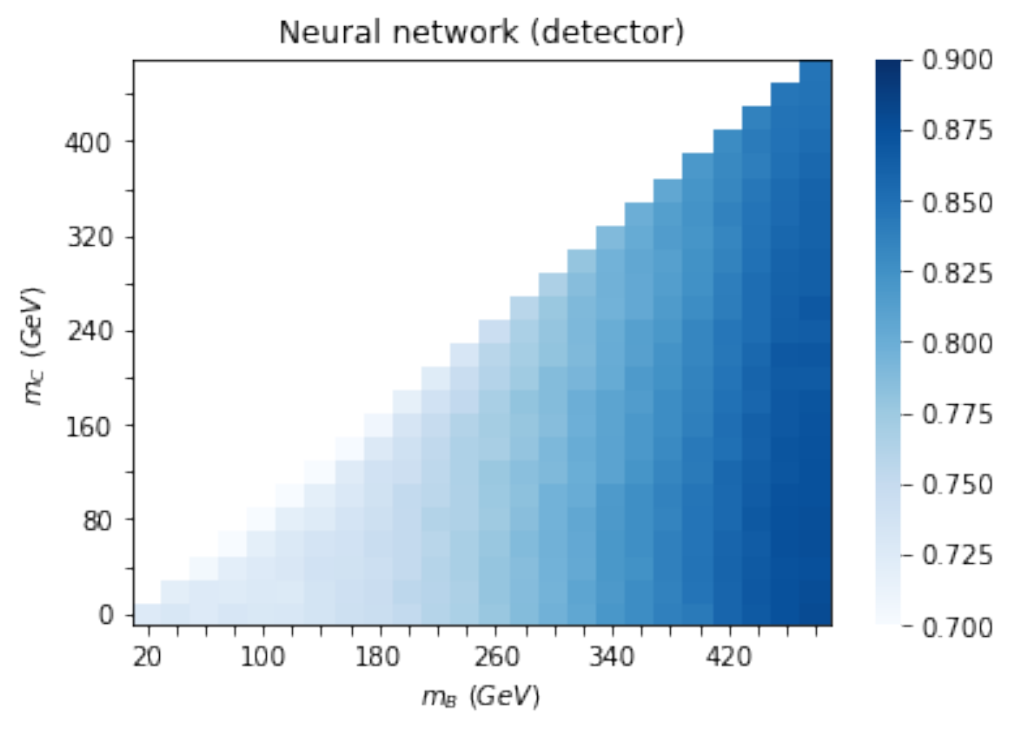} \\
\includegraphics[width=0.413\textwidth]{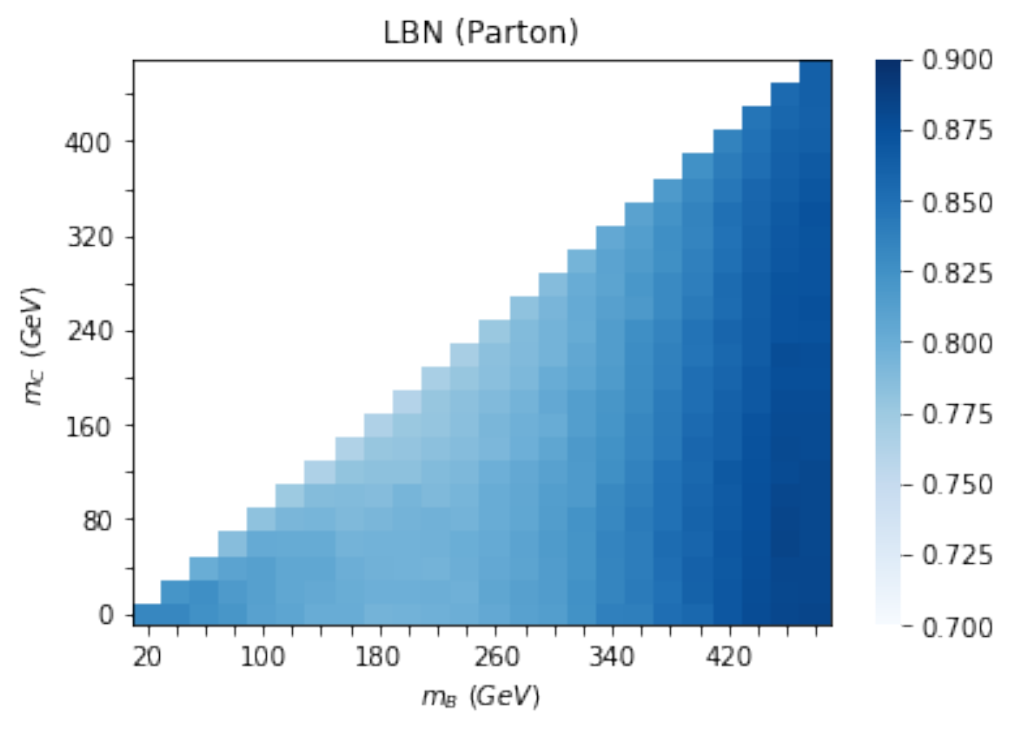}\hspace{0.5cm}
\includegraphics[width=0.413\textwidth]{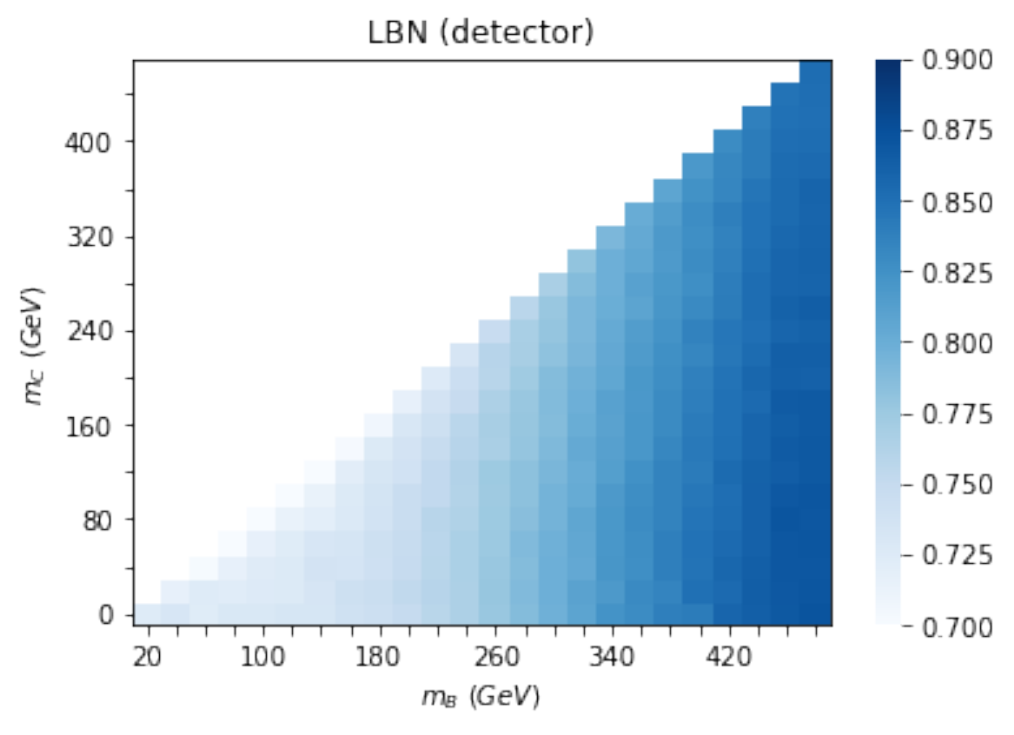}
\caption{Purity for choosing the correct partitioning with the endpoint method I (top), the hemisphere method (second row), DNN (third row) and LBN (bottom) in the two dimensional ($m_B$, $m_C$) mass space for $m_A = 500$ GeV and $m_B > m_C$. The purity in the left (right) panels are obtained for parton-level (detector-level) events. 
}
\label{fig:scan2d}
\end{center}  
\end{figure*}

\begin{figure*}[t!]
\begin{center}
\includegraphics[width=0.42\textwidth]{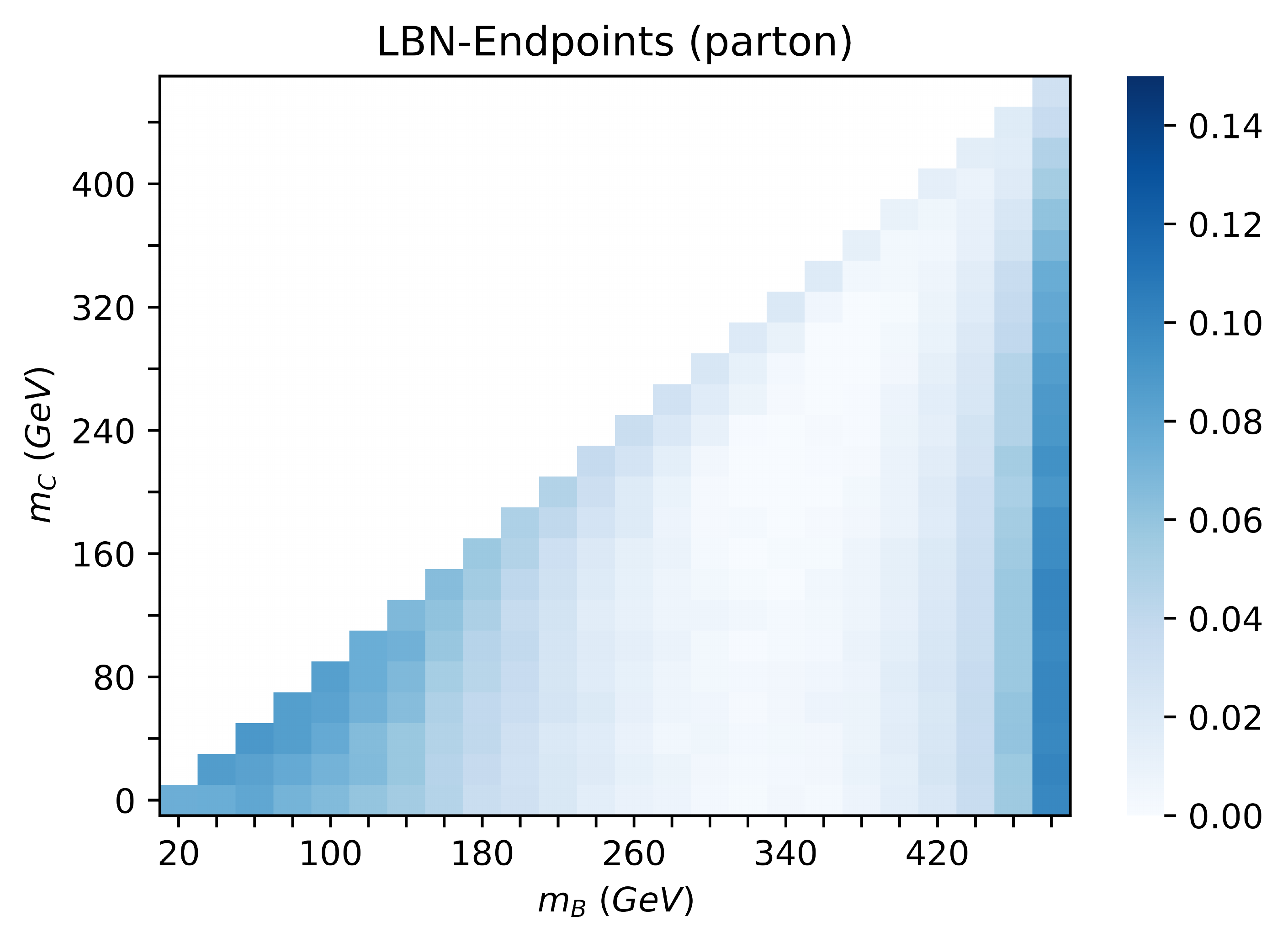} \hspace{0.5cm}
\includegraphics[width=0.42\textwidth]{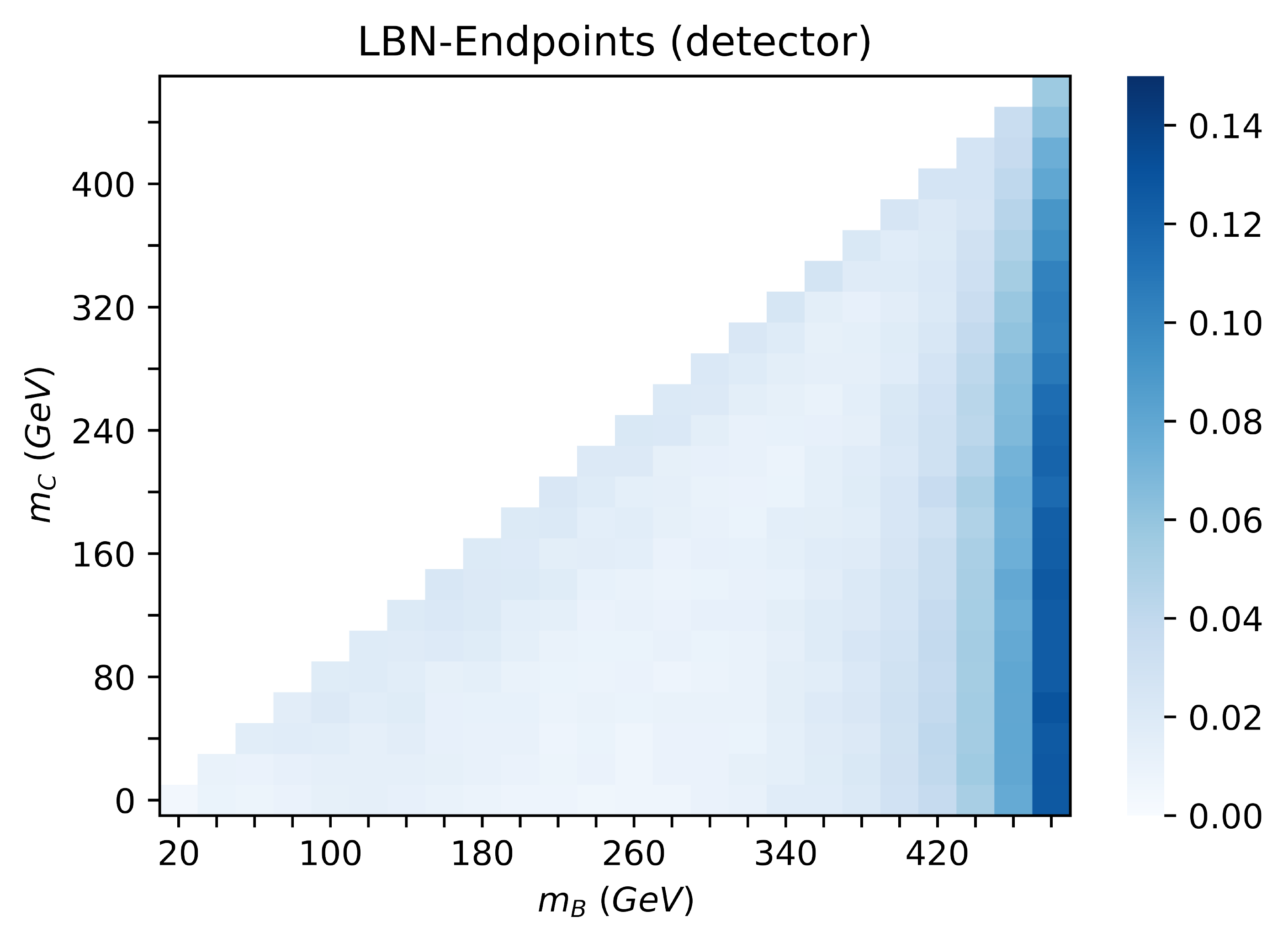} \\
\includegraphics[width=0.42\textwidth]{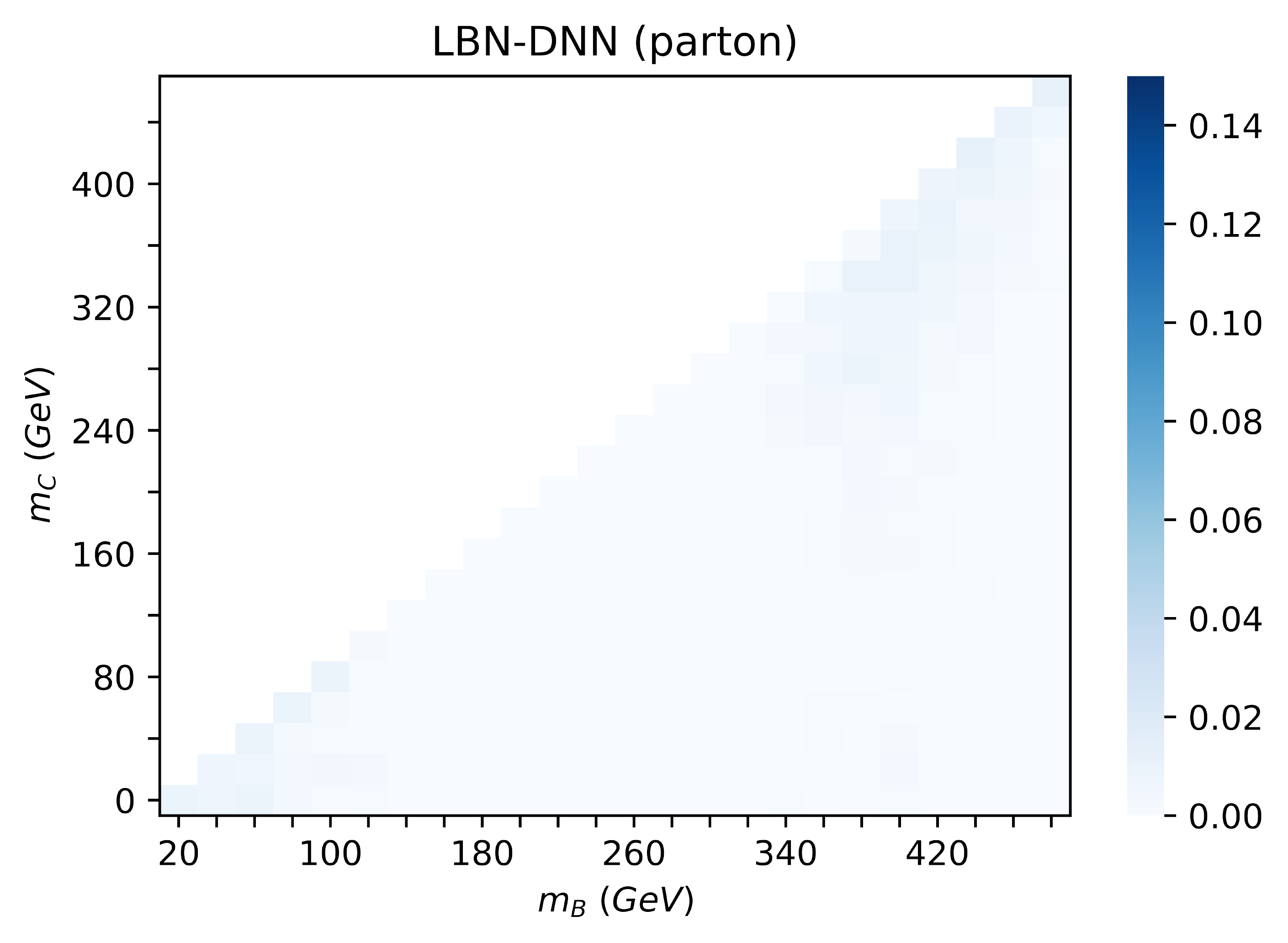} \hspace{0.5cm}
\includegraphics[width=0.42\textwidth]{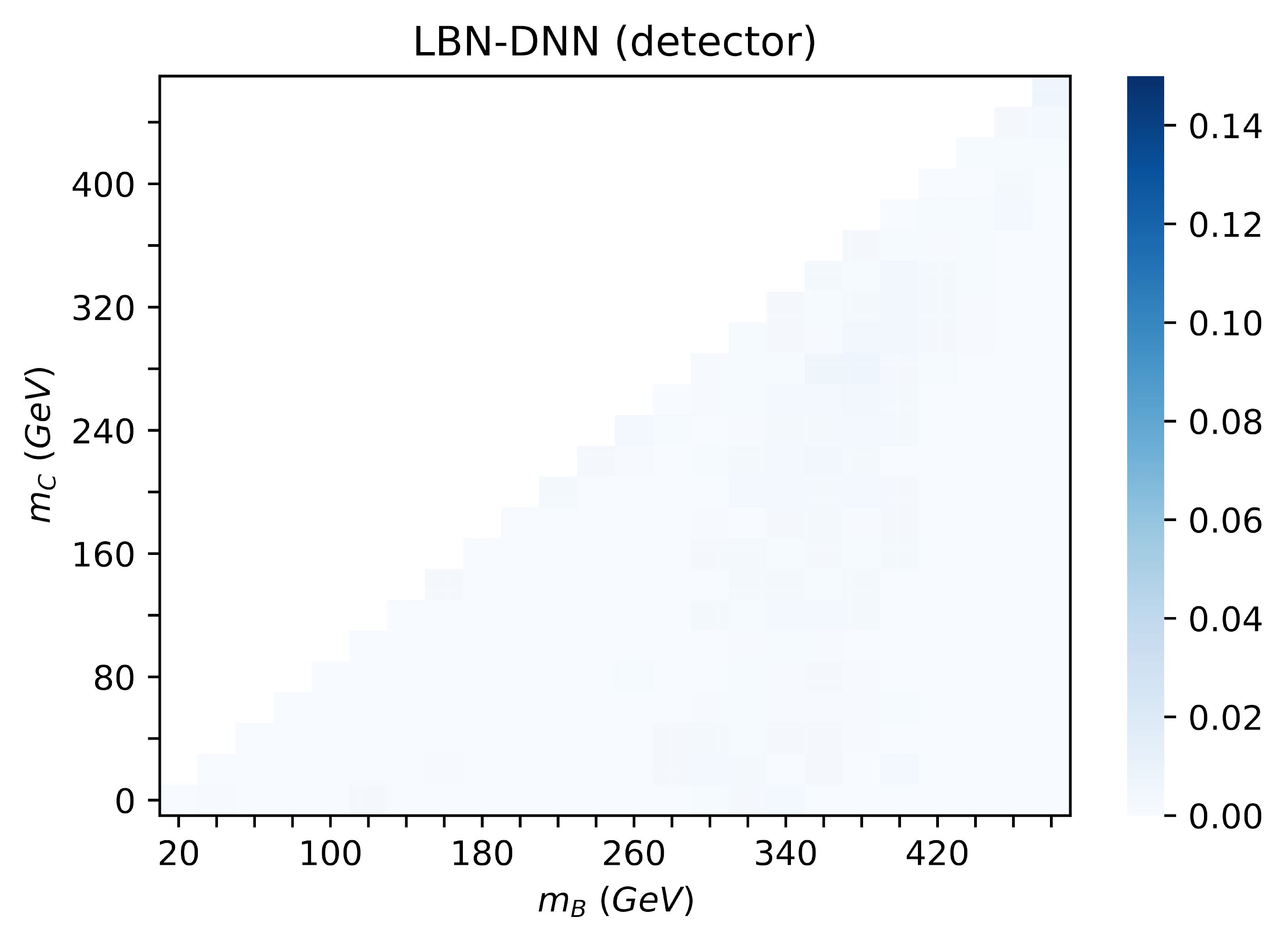}
\caption{
The purity-difference between LBN and endpoint method I (top) and between LBN and DNN (bottom) for parton-level (left) and detector-level (right) events.}
\label{fig:diff}
\end{center}  
\end{figure*}

For DNN and LBN, we randomly choose 100 mass points in $(m_B, m_C)$, covering the triangular parameter space. These points are a good representation of the two dimensional mass space that we are interested in. 
We prepare 100k events for each chosen mass point and mixed them all before feeding them into NN. Although we do not feed numerical values of the chosen masses, NN learns how to resolve the two-fold ambiguity by suitable interpolation in the entire two dimensional mass space \cite{Baldi:2016fzo,Kim:2021pcz}. 

Fig. \ref{fig:scan2d} summarizes the results of the four methods (endpoint method I in the first row, hemisphere method in the second row, DNN in the third row, and LBN in the last row) for the parton-level (left) and for the detector-level (right) events, respectively. 
In the case of the endpoint and hemisphere methods, our results for the parton-level events are consistent with those in Ref. \cite{Debnath:2017ktz}. We have added similar plots for detector-level events as well in the right column.

Analogous to Fig. \ref{fig:ROC} for the top quark decay, we see here that NNs outperform the traditional methods for the case of unknown mass spectrum as well. This is clearly depicted in Fig. \ref{fig:diff}, which shows the purity-difference between LBN and endpoint method I (top), and between LBN and DNN (bottom) for parton-level (left) and detector-level (right) events. LBN and DNN provide very similar performance in the wide range of the mass parameters (bottom panel), while LBN surpasses the endpoint method I (top panel), whose performance drops in the degenerate spectrum, $m_B \approx m_A$ (in the right side of the figure) or $m_B \approx m_C$ (along the diagonal), for both parton-level and detector-level, and $m_B \, , m_C \ll m_A$ (in the left-bottom corner) for parton-level.

\section{Discussion and outlook}
\label{sec:conclusion}

Resolving the combinatorial problem in collider experiments is crucial for the discovery of new physics and precision measurements. The simplest combinatorial problem is the two-fold ambiguity, which arises in the top quark pair production. As the most massive fundamental particle in the standard model, top quark is the only quark that decays before hadronization. 
The most precise measurements of top quark properties such as top quark mass are typically performed in the lepton-jet or dilepton channels. Therefore it is important to find the correct assignment for the reconstructed objects ($b$-tagged jet and a lepton), as the incorrect assignment will reduce the precision in the measurements of the top quark properties. 

In the dileptonic $t\Bar{t}$ events, we do not know the complete kinematic information due to the two neutrinos since they do not leave a trace in the detector. In particular, we do not know the individual momentum of each missing particle; we only know the total missing transverse momentum $\mptvec$. We are unable to reconstruct the missing momenta exactly, which poses the difficulty in assigning a lepton and a reconstructed $b$ quark pair.

In this paper, we have revisited with machine learning algorithms the combinatorial problem in the dileptonic $t\bar t$-like event topology. We have compared the performance of various algorithms against that of existing methods. In particular, we investigated the performance of attention-based networks, which has been found useful in the fully hadronic channel, and the Lorentz Boost Network, which is motivated by underlying physics principles. We found that most of the machine learning methods outperform the existing approaches based on kinematic variables. We then generalized the mass spectrum in consideration of new physics where the underlying mass spectrum is unknown, and therefore no kinematic endpoint information is available.
We showed that the purity for selecting the correct partition is greatly improved by utilizing the machine learning techniques, especially in the regions where particle spectrum is degenerate. 

A specific application of our study would be measurement of the top quark Yukawa coupling ($y_{t, {SM}}$) in the $t \bar t h$ production. As discussed in Refs. \cite{Goncalves:2018agy,Goncalves:2021dcu}, top quark reconstruction in the dilepton channel plays an important role. For example, Table \ref{table:summary} shows that the endpoints method gives 74.2\% efficiency and 87.4\% purity at detector-level (marked as {\color{cyan}\ding{72}} in Fig. \ref{fig:ROC}), which is comparable to results presented in Refs. \cite{Goncalves:2018agy,Goncalves:2021dcu}. 
Using the $M_2$ reconstruction and the sideband subtraction, the uncertainty ($\delta \kappa_t$) on the top quark Yukawa coupling ($\kappa_t = y_{t}/y_{t, {SM}}$, which is the deviation from the SM value) is calculated as $\delta \kappa_t \sim 0.096$ \cite{Goncalves:2021dcu} at the HL-LHC. 
On the other hand, our results show that the same purity (87.4\%) can be obtained at a higher efficiency using DNN (with 92\% efficiency) and LBN (with nearly 100\% efficiency), which would lead to a gain of 24\% and 35\% more events, respectively. Naive rescaling indicates that a gain $\alpha$ in the number of (both signal and background) events would lead to the $\alpha/2$ reduction in the precision on $\delta \kappa_t$. 
Following more accurate procedure in Ref. \cite{Goncalves:2021dcu}, we estimate the uncertainty to be reduced to $\delta \kappa_t \sim 0.086$ and $\delta \kappa_t \sim 0.082$ using DNN and LBN, respectively. 

Another application of our study would be testing Bell's inequality with top quark pair production \cite{Fabbrichesi:2021npl,Severi:2021cnj}. It is well known that the leptonic final state is maximally correlated to top quark polarization, which motivates the measurement of violation of Bell's inequality performed in the dilepton channel. One of main goals is to reconstruct the spin density matrix in the center of the momentum frame, which requires full reconstruction of two neutrinos, resolving the two-fold ambiguity. We anticipate that ML methods would help such a measurement \cite{Bell}.

Finally we would like to make brief comments on impacts of ISR/FSR and backgrounds, postponing detailed analysis in a future study. 
Our goal in this paper was to investigate different NN architecture in depth to resolve the combinatorial problem (two-fold ambiguity) without worrying about ISR/FSR and backgrounds so that we can try various ML methods and make fair comparison against existing approaches.
We took the ttbar dilepton production as a specific example, and then applied the ML methods to the case with arbitrary mass spectrum, keeping an application in BSM searches in mind. 
Therefore we intentionally ignored ISR/FSR and backgrounds, as their impacts are model-dependent in a sense that the hardness of ISR depends on the mass scale of new particles and backgrounds depends on the mass splitting. 

Although we have not investigated the effects of ISR/FSR using NNs, one can get a rough idea using the exiting methods for $t\bar t + X$ production. Refs. \cite{Goncalves:2021dcu,Goncalves:2018agy} studied the two-fold ambiguity in the $t\bar t h$ production with $h\to b\bar b$, where parton-shower and hadronization are simulated with PYTHIA. For example, Ref. \cite{Goncalves:2021dcu} finds the efficiency of the endpoint II to be 78\%, while our detector-level efficiency in this paper is 74\%. The small difference is due to different set of cuts in data preparation. In Refs. \cite{Goncalves:2021dcu,Goncalves:2018agy}, two hardest $b$-tagged jets are chosen, which effectively rejects contamination arising from ISR/FSR, resulting in similar performance in the absence ISR/FSR. 

Effect of ISR for top quark production is studied in details in Ref. \cite{Baringer:2011nh}, using $M_{T2}$ and $m_{b\ell}$ method, where ISR/FSR and hadronization, and detectors effects are simulated with PYTHIA and PGS, respectively. Taking two leading $b$-tagged jets as $b$-quark candidates from the top decay, they find an efficiency of 51.7\% with a purity of 94.9\% at the 7 TeV LHC, which is comparable to the endpoint I in the right panel of Fig. \ref{fig:ROC}. This comparison indicates that effects of ISR/FSR and hadronization are mild or negligible when resolving the two fold ambiguity for the dilepton top quark production. 

Effects of ISR in reconstruction of new particle masses for new physics processes have been studied in Ref. \cite{Alwall:2009zu}. They developed a novel technique to reduce ISR effects, taking gluino pair production and its decay to two jets and neutralino, $\tilde g \tilde g + j \to 5 \, {\rm jets} + \tilde\chi_1^0\tilde\chi_1^0$, as an example.  With 5 jets candidates, they compute $M_{T2}(i)$ excluding the $i$-th jet ($i=1,\cdots,5$) and define $M_{T2}^{\rm min} = \min\limits_{i=1,\cdots,5} \Big ( M_{T2}(i) \Big )$. Therefore by construction, $M_{T2}^{\rm min} < M_{T2}^{\rm endpoint}$, and the $i_{\rm min}$-th jet that satisfies $M_{T2}(i_{\rm min}) = M_{T2}^{\rm min}$ is considered to be the ISR jet. Surprisingly, this simple kinematic method gives reasonably good efficiency in identifying ISR jet. For a given mass spectrum ($m_{\tilde g}=685$ GeV and $m_{\tilde\chi_1^0}=101.7$ GeV), they identify the ISR jet among 5 jets correctly 29\% of time. This number increases up to 44\% with a cut $\min(M_{T2}) > 500$ GeV. A similar method is used to distinguish $t\bar t$ and $tW$ production \cite{Kim:2015uea}. 

Similarly the importance of backgrounds also depends on the details of new physics including mass spectrum. For example, one can consider a new physics scenario, where the mass difference between $B$ and $C$ is much larger than $m_W$ ($m_B - m_C \gg m_W$), which would lead to two very high $p_T$ leptons with little backgrounds. On the other hand, if $m_B - m_C \lesssim m_W$, we would suffer from the $t\bar t$ background. Therefore to better estimate their impacts, it is appropriate to consider a specific new physics case with a fixed mass spectrum. 

Even within SM, the effects of ISR/FSR and backgrounds depend on what other particles are produced along the two top quarks. Note that all discussion in this paper is valid for any $t\bar t + X$ processes, as long as $X$ is reconstructable.  For example, one can consider $t\bar t h$ with $h \to \gamma \gamma$ or $t \bar t h$ with $h \to b \bar b$, where dominant backgrounds are different. Specifically, for top quark production, ISR/FSR should be relatively harmless for our problem given the high $b$-tagging efficiency and small fake rates expected at the HL-LHC \cite{CERN-LHCC-2017-021,Baringer:2011nh}. Therefore in the case of $t\bar t$ production, background is not really a big concern. One can further reduce the backgrounds at the cost of statistics, utilizing various kinematic methods, as those methods themselves are optimized for signal ($t\bar t$) and backgrounds processes would violate the endpoint structures \cite{Goncalves:2021dcu,Goncalves:2018agy}. 
An interesting question now is how well one can improve these results using NNs, the details of which we reserve for a future study.

So far we have focused on resolving the two-fold ambiguity. One of advantage of using the kinematic methods is their byproduct, i.e., some methods provide ansatz for the momentum of the missing particles, which can be used to reconstruct the full final state approximately. A similar study with neural networks has been done utilizing Lorentz structure of the four momenta \cite{LBN}. Generalizing such a method for arbitrary mass spectrum would be useful in search for new physics beyond the standard model. We expect that NN-inspired reconstruction will help to expedite discovery as well as the precision measurement.

\bigskip
\emph{Acknowledgements:} 
We thank Myeonghun Park for useful discussion and Doojin Kim for helpful comments on the manuscript.
CD is supported in part by the US DOE under grant No. DE-SC0019474, and in part by the State of Kansas EPSCoR grant program.
LH is supported by the Fundamental Research Funds for the Central Universities and the Bureau of International Cooperation, Chinese Academy of Sciences. 
JK was supported in part by the National Research Foundation of Korea (NRF) grant funded by the Korea government (MSIT) (No. 2021R1C1C1005076), and in part by the international cooperation program managed by the National Research Foundation of Korea (No. 2022K2A9A2A15000153, FY2022).
KK acknowledges support from the US DOE, Office of Science under contract DE-SC0021447 and the University of Kansas General Research Fund allocation. 
The work of DS was supported by the US Department of Energy under grant DE-SC0010008.

\appendix

\section{Overview of existing methods}
\label{sec:existingmethods}
\begin{figure*}[t!]
\begin{center}
\includegraphics[width=0.43\textwidth]{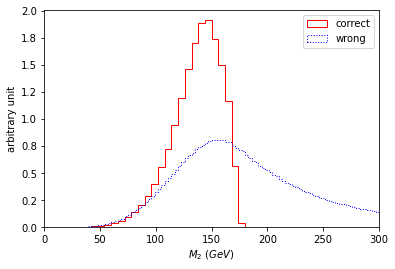} \hspace*{0.5cm}
\includegraphics[width=0.43\textwidth]{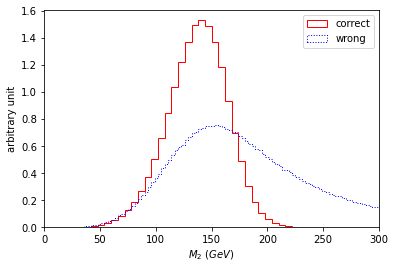}
\caption{$M_{2}$ distribution of the correct (red, solid) and incorrect (blue, dotted) pair for parton-level events (left) and detector-level events (right).}
\label{fig:m2}
\end{center}  
\end{figure*}

There are several methods to resolve the two-fold ambiguity in the dilepton $t\bar t$ production.
Although each method is introduced elsewhere, we did not find detailed description on the topic. 
Moreover there is no comprehensive study on the two-fold ambiguity, which compares performance of all different methods using the same set of events. We find it valuable to provide a brief review on various existing methods in this Appendix, including performance of each method, and proper comparison. 
Although we report the purity and efficiency of individual method in resolving the two fold-ambiguity, a method with a higher efficiency or a higher purity does not necessarily mean a better method. Each method is unique, has different motivations and is based on different set of assumptions. 
Depending on the underlying event topology and the target study point, they
may show different levels of performance, hence it is prudent to keep as many tools as possible in the analysis toolbox. 

\subsection{Endpoint methods}
\label{sec:endpoint}

For events with two missing particles, the on-shell constrained $M_2$ variable \cite{Cho:2014naa,Kim:2017awi} provides a good estimate for the unmeasured invisible momenta and thus can be useful to discriminate combinatorial ambiguities~\cite{Barr:2011xt,Debnath:2017ktz,Kim:2017awi}. It is defined as a $(3+1)$-dimensional version of $M_{T2}$ \cite{Lester:1999tx,Barr:2003rg,Burns:2008va,Konar:2009qr}:
\begin{align}
M_{2} (\tilde m) &\equiv \min_{\vec{q}_{1},\vec{q}_{2}}\left\{\max\left[M_{P_1}(\vec{q}_{1},\tilde m),\;M_{P_2} (\vec{q}_{2},\tilde m)\right] \right\} ,
\nonumber\\
\mptvec &=\vec{q}_{1T}+\vec{q}_{2T}  \;,
\label{eq:m2def}
\end{align}
 where the {\em actual} parent masses, $M_{P_i}$, are considered instead of their transverse masses, $M_{T P_i}$ ($i=1, \, 2$). 
 The $\tilde{m}$ is the test mass, which we take to be zero in our study.
Note that the minimization is performed over the 3-component momentum vectors $\vec{q}_{1}$ and $\vec{q}_{2}$ of the two missing particles~\cite{Barr:2011xt}, assuming the missing transverse momentum constraint as shown in Eq. (\ref{eq:m2def}).
At this point $M_{T2}$ and $M_2$ are known to be equivalent, in the sense that the resulting two variables will have the same numerical value $M_2 = M_{T2} \leqslant  \max(M_{P_1}, M_{P_2})$~\cite{Ross:2007rm,Barr:2011xt,Cho:2014naa}. 
Fig. \ref{fig:m2} shows $M_{2}$ distribution of the correct (red, solid) and incorrect (blue, dotted) pair for parton-level events (left) and detector-level events (right). The correct pairing respects the mass bound, while the incorrect paring goes beyond the expected endpoint, which is the mass of the top quark in this example.

However, for the $t \bar t$ production considered in this paper (in general, $t\bar t + X$, where $X$ is fully reconstructed), the value of the $W$-boson mass $m_W$ is experimentally known, and therefore we can introduce the following variable in the ($b\ell$) subsystem:
\bea
M_{2CW}^{(b\ell)} &\equiv& \min_{\vec{q}_{1},\vec{q}_{2}}\left\{\max\left[M_{t_1}(\vec{q}_{1},\tilde m),\;M_{t_2} (\vec{q}_{2},\tilde m)\right] \right\},\label{eq:m2CWdef} \nonumber\\
\mptvec&=&\vec{q}_{1T}+\vec{q}_{2T}   \,, \\
M_{t_1}&=& M_{t_2}\,, \nonumber  \\
M_{W_1}&=& M_{W_2} = m_W \,.\nonumber
\eea
Here the second constraint $M_{t_1}= M_{t_2}$ requires the equality of two parent mass without use of a specific numerical value, while the true $W$ mass is used in the third constraint $M_{W_1}= M_{W_2} = m_W$.
Similarly, taking the top quark mass $m_t$ in the minimization, we can define a new variable in the ($\ell$) subsystem:
\bea
M_{2Ct}^{(\ell)} &\equiv& \min_{\vec{q}_{1},\vec{q}_{2}}\left\{\max\left[M_{W_1}(\vec{q}_{1},\tilde m),\;M_{W_2} (\vec{q}_{2},\tilde m)\right] \right\},\nonumber\\
\mptvec&=&\vec{q}_{1T}+\vec{q}_{2T}   \,,  \label{eq:m2Ctdef} \\
M_{W_1}&=& M_{W_2} \,, \nonumber  \\
M_{t_1}&=& M_{t_2} = m_t \,.\nonumber 
\eea
By construction, $M_{2Ct}^{(\ell)} \leqslant  m_W$ and $M_{2CW}^{(b\ell)} \leqslant  m_t$. 

\begin{figure*}[t]
\begin{center}
\includegraphics[width=0.43\textwidth]{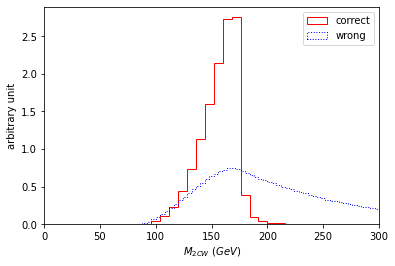} \hspace*{0.5cm}
\includegraphics[width=0.43\textwidth]{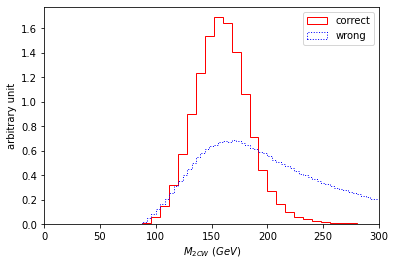} \\
\includegraphics[width=0.43\textwidth]{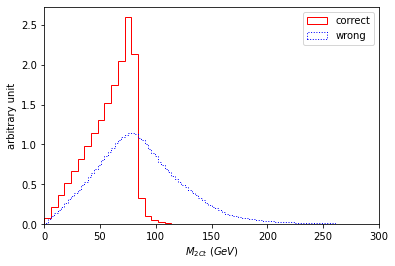}\hspace*{0.5cm}
\includegraphics[width=0.43\textwidth]{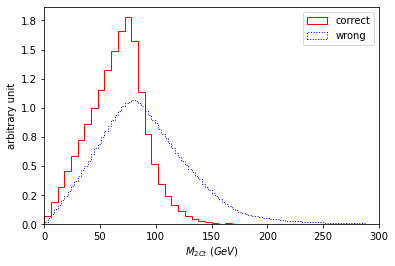}
\caption{$M_{2CW}$ (top, Eq. (\ref{eq:m2CWdef})) and $M_{2Ct}$ (bottom, Eq. (\ref{eq:m2Ctdef})) distributions of the correct (red, solid) and incorrect (blue, dotted) pair for parton-level events (left) and detector-level events (right).}
\label{fig:m2cw_m2ct}
\end{center}  
\end{figure*}

Fig. \ref{fig:m2cw_m2ct} shows $M_{2CW}$ (top) and $M_{2Ct}$ (bottom) distributions of the correct (red, solid) and incorrect (blue, dotted) pair for parton-level events (left) and detector-level events (right), respectively. We note that $M_{2CW}$ shows a sharper distribution compared to $M_2$ distribution, while both respect the same endpoint (top quark mass). For both cases, the input mass (or test mass) is zero. However, $M_{2CW}$ distrbiution begins from $m_W$ since the numerical value of the $W$-boson mass is imposed during the minimization as shown in Eq. (\ref{eq:m2CWdef}).

For the minimization of $M_{2}$, $M_{2Ct}$ and $M_{2CW}$, we use OPTIMASS~\cite{Cho:2015laa}.
While these mass-constraining variables are proposed for mass measurement originally,
one could use them for other purposes such as measurement of spins and couplings \cite{Baringer:2011nh,Debnath:2017ktz}. In our study, we use these variables to fully reconstruct the final state of our interest, with the unknown neutrino momenta obtained via minimization procedure. These momenta may or may not be true particle momenta but they provide important non-trivial correlations with other visible particles in the final state, which help reconstruction.

\begin{figure*}[t]
\begin{center}
\includegraphics[width=0.43\textwidth]{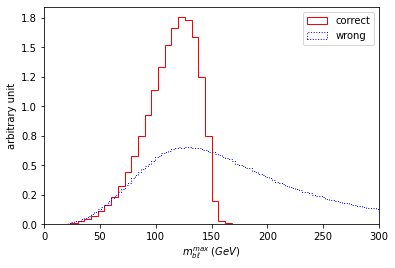}\hspace*{0.5cm}
\includegraphics[width=0.43\textwidth]{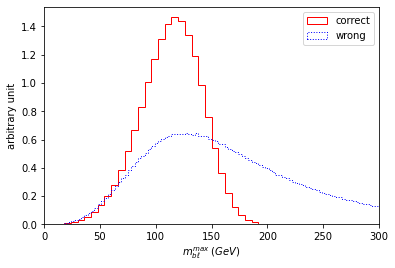}
\caption{$max\{ m_{b\ell}^{(i)} \}$ distribution of the correct (red, solid) and incorrect (blue, dotted) pair for parton-level events (left) and detector-level events (right).}
\label{fig:mbl}
\end{center}  
\end{figure*}

The other useful kinematic variable is the invariant mass $m^{(i)}_{b\ell}$ of $b$ and $\ell$ in $i$-th pairing ($i=1,2$). The invariant mass distribution for the correct pairing is bounded by the minimum and maximum values ($m_{b\ell}^{min} \leqslant  m_{b\ell} \leqslant  m_{b\ell}^{max}$), which are given by 
\begin{eqnarray}
\hspace*{-0.25cm}\left ( m_{b\ell}^{max/min} \right)^2 &=&
\frac{1}{2} \Big ( m_t^2-m_W^2 + m_b^2   \label{eq:mbl} \\
&& \hspace*{-2cm}  \pm \sqrt{\big( ( m_t- m_b)^2-m_W^2\big)
   \big (( m_t +m_b)^2- m_W^2\big )} \Big ) \, ,\nonumber 
\end{eqnarray}
which become $m_{b\ell}^{max}= \sqrt{ m_t^2 - m_W^2}$ and $m_{b\ell}^{min}=0$ in the $m_b \to 0$ limit.
Fig. \ref{fig:mbl} shows the $m_{b\ell}$ distribution of the correct (red, solid) and incorrect (blue, dotted) pair for parton-level events (left) and detector-level events (right). We take the larger ($max\{ m_{b\ell}^{(i)} \}$) of two possible invariant masses for each choice of partitioning.
\begin{figure*}[t!]
\begin{center}
\includegraphics[width=0.48\textwidth]{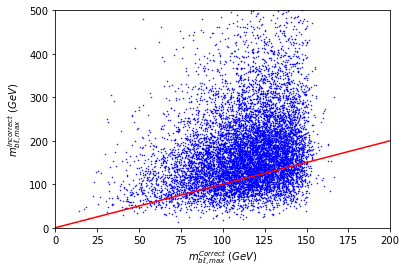} \hspace*{-0.1cm}
\includegraphics[width=0.48\textwidth]{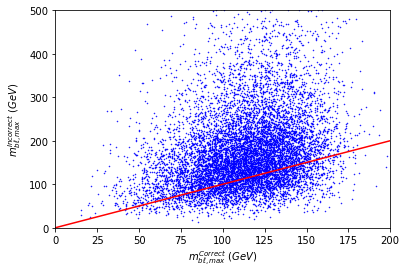}
\caption{Scatter distribution of ($m_{b\ell}^{correct}$, $m_{b\ell}^{incorrect}$) for parton-level events (left) and detector-level events (right). The red-diagonal line represents $m_{b\ell}^{correct}=m_{b\ell}^{incorrect}$.}
\label{fig:mblscatter}
\end{center}  
\end{figure*}

Now we follow the procedure described in Ref.~\cite{Debnath:2017ktz} to resolve the two-fold ambiguity. 
For each event, we compute the following 3-dimensional vector for both correct and incorrect pairings:
\begin{equation}
\Big(   \,
m_{b\ell}^{max}-\max_i\{m^{(i)}_{b\ell}\}, \,
m_t - M_{2CW}^{(b\ell)}, \,
m_W - M_{2Ct}^{(\ell)} \,
\Big) .
\label{setofthreeWt}
\end{equation}
The correct combination would respect the anticipated endpoints of $m_{b\ell}$, $M_{2CW}^{(b\ell)}$ and $M_{2Ct}^{(\ell)}$, leading to positive components of above 3-dimensional mass space. On the other hand, the incorrect pairing could give either sign. Therefore by requiring that the partition which gives more ``plus" signs as the ``correct" one, we can resolve the two-fold ambiguity. Then, we treat the corresponding momenta of the two missing particles, which are obtained via the minimization procedure, as ``approximate'' momenta of the two missing neutrinos. If both partitions give the same numbers of positive and negative signs, we discard such events, since they are  ``unresolved cases''. 
For parton-level events, we find such unresolved events are 23\%, while it is 26\% for detector-level events. 

With the definitions of efficiency and purity in Eqs. (\ref{eq:eff})-(\ref{eq:purity}), 
selecting the resolved events only, the endpoint method leads to 
77\% efficiency and 96\% purity for parton-level events, and 
74\% efficiency and 87\% purity  for detector-level events, respectively. 
We call this method as the endpoint method II, which is shown as ({\color{cyan}\ding{72}}) in Table \ref{table:summary} and Fig. \ref{fig:ROC}. 
Similar methods were considered in the literature, using ($m_{b\ell}$, $p_t$) \cite{Rajaraman:2010hy}, or ($m_{b\ell}$, $M_{T2}$)  \cite{Baringer:2011nh,Choi:2011ys}. We find that the latest study with ($m_{b\ell}$, $M_{2Ct}$, $M_{2CW}$) in Ref. \cite{Debnath:2017ktz} gives the best result concerning the combinatorial problem. 

Although the use of mass spectrum resolves the two-fold ambiguity more accurately, it is instructive, and perhaps necessary in some examples, to repeat a similar analysis without using mass information explicitly. We use \big ($m_{b\ell}$, $M_{2CC}^{(b\ell)}$, $M_{2CC}^{(\ell)}$\big ) without prior knowledge of mass spectrum. 
For each event, there are two possible values for each component of \big ($m_{b\ell}$, $M_{2CC}^{(b\ell)}$, $M_{2CC}^{(\ell)}$\big ). Choosing the combination that gives more smaller components as the correct one, we obtain 81.6\% purity for parton-level events, and 78.9\% purity for detector-level events, respectively. Since there are three quantities that we compare, in this case there is no unresolved event and therefore the efficiency is 100\%. This is denoted as the endpoint method I in Table \ref{table:summary} and Fig. \ref{fig:ROC}. 

The endpoint method is very general and can be extended easily to different event topologies.  
However, it also has a few issues. First, the endpoints are sensitive to the detector effects and get smeared significantly. A proper de-convolution procedure (using the transfer function) is required  for a better performance. Secondly, the finite widths of intermediate particles (top quark and $W$ in this case) also affect the shape of the kinematic distributions, in which case, the correct pairing could violate the expected endpoint. Finally there are quite large number of unresolved events, which are discarded, when mass information is imposed. These issues motivate us to explore different methods to maximize both efficiency and purity.   

\subsection{Hemisphere method and recursive jigsaw reconstruction}
\label{sec:hemisphere}

Partitioning reconstructed particles into  two decay chains
is often addressed by the so-called ``hemisphere" algorithm, developed originally within CMS \cite{Ball:2007zza} and later adopted in many phenomenological studies~\cite{Matsumoto:2006ws,Cho:2007dh,Nojiri:2008hy}. Using the standard hemisphere method, we cluster the visible particles into two groups by keeping the invariant mass of each group to a minimum. For the $t\bar t$-like topology considered in this paper, it is straightforward to see that the hemisphere method is nothing but a variation of invariant mass method without relying on the endpoint. The pairing whose invariant mass is smaller is chosen to be the correct pair and the other is chosen to be the incorrect one. Since the hemisphere method does not use the numerical value of the endpoint, there is no violation of endpoint and therefore we do not discard any event, keeping 100\% efficiency. We obtain the 78\% purity for parton-level events and 77\% purity for detector-level events. These events are ones in the left-upper corner of the $m_{b\ell}^{correct}=m_{b\ell}^{incorrect}$ line (red, solid) in Fig. \ref{fig:mblscatter}.
An advantage of this method is that one could obtain relatively good purity with 100\% efficiency without using the mass spectrum via fast computation. The results are not very sensitive to the detector effects. 
However, if a high purity sample is required, the method must be extended at the cost of statistics. For example, as we discussed in the previous method, one can further improve on the hemisphere algorithm by suitable cuts on the invariant mass and either the jet $p_T$ \cite{Rajaraman:2010hy} or $M_{T2}$ \cite{Baringer:2011nh,Choi:2011ys}.

A similar idea is discussed in the 
Recursive Jigsaw Reconstruction method. It is a technique for analyzing reconstructed particles in the presence of kinematic unknowns arising from the unmeasured particles, and the combinatoric unknowns associated with indistinguishable particles, respectively \cite{Jackson:2017gcy}.
The method provides a very general framework, which can be applied to various processes at collider experiments. In particular, the dileptonic $t\bar t$ production is one of the examples discussed in Ref. \cite{Jackson:2017gcy}. Due to the simple nature of the two-fold ambiguity, the Recursive Jigsaw Reconstruction method becomes very similar to the hemisphere method. It takes the smaller of the two squared mass sum, $ \min \left ( m_{b_1 \ell^+}^2 + m_{b_2\ell^-}^2, m_{b_2 \ell^+}^2 + m_{b_1\ell^-}^2 \right )$, as the correct pair.
In other words, it chooses the  combination where the sum of four vector inner products is smallest, effectively pairing particles flying closer together as expected from a common decay source.
This algorithm gives 
76.2\% of purity for parton-level events and 75.7\% for detector-level events, respectively. We also used {\tt RestFrames} \cite{Jackson:2017gcy,restframes} and obtained a similar purity. These results are very comparable to what we have obtained using the hemisphere method. 

\begin{figure*}[t!]
\begin{center}
\includegraphics[width=0.45\textwidth]{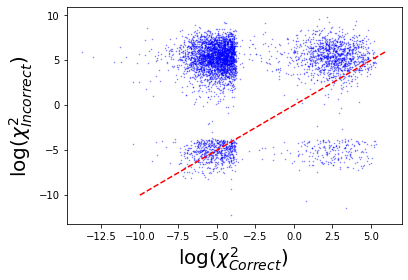}\hspace*{0.2cm}
\includegraphics[width=0.45\textwidth]{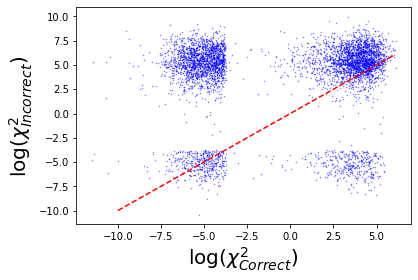}
\caption{$\chi^2$ value of the correct ($x$-axis) and the incorrect pair ($y$-axis) for parton-level events (left) and detector-level events (right). The red-diagonal line represents $\chi^2_{12}=\chi^2_{21}$.} 
\label{fig:topness}
\end{center}  
\end{figure*}

\subsection{Topness}
\label{sec:topness}

The topness ($T$) is originally proposed in search for supersymmetry \cite{Graesser:2012qy}, and then later a modified version is used in search for the double Higgs production \cite{Kim:2019wns,Kim:2018cxf}. It is nothing but a simple mass fitting in the $t\bar t$-like topology with a given mass spectrum. 

The definition of the $\chi^2$ statistic is given by 
\begin{align}
    \chi^2_{ij} \equiv \min_{\mptvec= \vec p_{\nu T}+\vec p_{\Bar{\nu}T}} & \left [\dfrac{\big( m^2_{b_i \ell^+ \nu}-m^2_t\big)^2}{\sigma^4_t}+\dfrac{\big( m^2_{\ell^+ \nu}-m^2_W\big)^2}{\sigma^4_W}    \right. \nonumber \\
    &\hspace*{-1.5cm}+ \left. \dfrac{\big( m^2_{b_j \ell^- \Bar{\nu}}-m^2_t\big)^2}{\sigma^4_t}+\dfrac{\big( m^2_{\ell^- \Bar{\nu}}-m^2_W\big)^2}{\sigma^4_W}\right ] \, ,  
\end{align}
where $b_1$ and $b_2$ are the $p_T$-ordered $b$-tagged jets.
The topness $T$ is defined as $T = \min (\chi^2_{12}, \chi^2_{21})$. We find the momentum information of the two neutrinos via minimization algorithm over the missing transverse momentum constraint, $\mptvec= \vec p_{\nu T}+\vec p_{\Bar{\nu}T}$. 
Then the correct pair is identified as the combination of $b$ and $\ell$, which gives the the smaller of the two $\chi^2$ value, being more consistent with the top quark and $W$-boson masses.  The correctly identified events reside in the left-upper corner of the diagonal line in Fig. \ref{fig:topness}. We  obtain the 85\% purity for parton-level events and 81\% purity for detector-level events, keeping 100\% efficiency, as shown as ({\color{asparagus}\ding{117}}) Table \ref{table:summary} and Fig. \ref{fig:ROC}. Note that in our minimization, we choose $\sigma_t =\sigma_W = 5$ GeV following Refs. \cite{Kim:2019wns,Kim:2018cxf}. We find that the purity is not very sensitive to the choice of the $\sigma$ parameters.

\subsection{Kinematic likelihood fitter}
\label{sec:KLfitter}

The Kinematic Likelihood Fitter (KLFitter)\footnote{The source code can be found from https://github.com/KLFitter/KLFitter.} \cite{Erdmann:2013rxa} is a library for kinematic fitting using a likelihood approach developed for the top quark reconstruction.
The reconstruction of dileptonic $t \bar t$ events utilizes the neutrino-weighting method to solve the under-constrained kinematic system in the final state with two $b$-tagged jets,  two charged leptons and the missing transverse momentum. 
The likelihood consists of three parts as shown in Eq.  (\ref{eq:KLFitter}). 
\begin{eqnarray}
    {\cal L} = &&\prod_{i=x,y} {\cal G}\left(E^{miss}_{i}|p^{\nu_1}_i, p^{\nu_2}_i, \sigma^{miss}_i\left(m_t, m_W, \eta_{\nu_1}, \eta_{\nu_2}\right  )\right)\nonumber \\
    & \times&  \prod_{i=1}^2 {\cal G}\left(\eta_{\nu_i}|m_t\right)\times\left(m_{\ell_1,q_1}+m_{\ell_2,q2}\right)^{\alpha} \label{eq:KLFitter}\\
    &\times&  \prod_{i=1}^2 
    {\cal W}_{jet}\left(p_{jet, i}^{\; detector}|p_{jet,i}^{\; parton} \right)  \nonumber\\
    & \times&  \prod_{i=1}^2 {\cal W}_{\ell}\left( p_{\ell, i}^{\; detector}| p_{\ell, i}^{\; parton}\right)\nonumber \, .
\end{eqnarray}
The Gaussian distribution ${\cal G}(\cdots)$ in the first line of Eq. (\ref{eq:KLFitter}) is two-dimensional and attempts to fix the neutrino momenta via the missing transverse momentum constraint.
The second line contains two one-dimensional Gaussian distributions multiplied by the inverse of the invariant masses. 
$\alpha$ is a tuning parameter of the likelihood and we use the default value $\alpha=-2$. This choice is consistent with choosing the smaller value of invariant masses, which could increase the likelihood.
The last two lines include the transfer functions for the two charged leptons (${\cal W}_\ell$) and the two jets (${\cal W}_{jet}$), which are defined in Eq. (\ref{eq:transfer1}) and Eq. (\ref{eq:transfer2}), respectively.
The transfer function contains the response of the detector, and is the conditional probability to observe a detector-level event for a given parton-level configuration.  We use the transfer function introduced in Refs. \cite{Artoisenet:2010cn,Artoisenet:2008zz,talkmem,D0:2011fla}, 
\begin{eqnarray}
&&\hspace*{-1.1cm}{\cal W}_{\ell}\left(p_{\ell,i}^{\; detector}| p_{\ell,i}^{\; parton}\right) = 1 
~~{\rm for~}p_{\ell,i}^{\; detector} = p_{\ell,i}^{\; parton}  ,
\label{eq:transfer1} \\
&&\hspace*{-1.1cm}{\cal W}_{jet}\left(p_{jet,i}^{\; detector}|p_{jet,i}^{\; parton} \right) =  \frac{1}{\sqrt{2\pi} \, (f_2 + f_3 f_5)} \nonumber \\
&&  \hspace*{1cm}\times \Big ( \exp^{-\frac{(\Delta-f_1)^2}{2 f_2^2}} +\, f_3 \, \exp^{-\frac{(\Delta-f_4)^2}{2 f_5^2}} \Big ) \, ,  \label{eq:transfer2}
\end{eqnarray}
where $\Delta = p_{jet,i}^{\; detector} - p_{jet,i}^{\; parton}$ and $f_j = a_j + p_{jet,i}^{\; parton} b_j$. The parameters $a_j$ and $b_j$ are determined from fully simulated $t \bar{t}$ events which are given in Ref. \cite{Artoisenet:2010cn,Artoisenet:2008zz,talkmem}. 
We use ${\cal W}_{\ell}\left(p_{\ell,i}^{\; detector}| p_{\ell,i}^{\; parton}\right) = 1$ for 
$p_{\ell,i}^{\; detector}= p_{\ell,i}^{\; parton}$ (otherwise ${\cal W}_{\ell}=0$) and ${\cal W}_{jet}\left(p_{jet,i}^{\; detector}|p_{jet,i}^{\; parton} \right) = 1$ for 
$p_{jet,i}^{\; detector}= p_{jet,i}^{\; parton} $ (otherwise ${\cal W}_{jet}=0$) for the parton-level analysis. 
 The likelihood distinguishes between neutrinos and anti-neutrinos, and the charged leptons are paired accordingly. We refer to Ref. \cite{Erdmann:2013rxa} for further details.
\begin{figure}[t!]
\begin{center}
\includegraphics[width=0.46\textwidth]{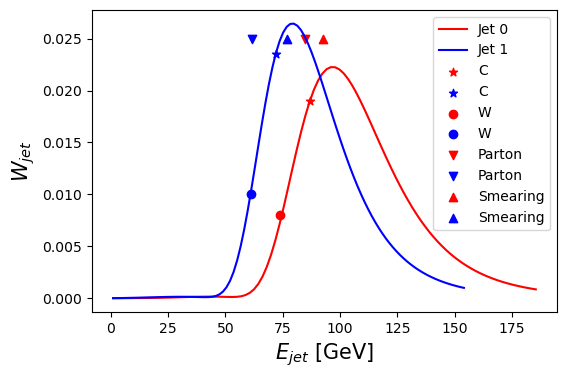}
\caption{
Distribution of the transfer function as a function of $E_{jet}$ for the two $b$-quark candidates in a sample event.
True MC inputs are marked as \ding{116} for parton-level energy (without detector effects) and as \ding{115} for smeared energy (with detector effects), respectively. The likelihood method fixes the most probably energy of jets, which are denoted by \ding{72} for correct pairing and \ding{108} for incorrect paring, respectively. 
\label{fig:trasfer} }
\end{center}  
\end{figure}

Fig. \ref{fig:trasfer} illustrate how above procedure works. 
Two solid curves represent the distribution of the transfer function as a function of $E_{jet}$ for the two $b$-quark candidates in a sample event. True MC inputs are marked as \ding{116} for parton-level energy (without detector effects) and as \ding{115} for smeared energy (with detector effects), respectively. The likelihood method fixes the most probably energy of jets, which are denoted by \ding{72} for correct pairing and \ding{108} for incorrect paring, respectively. Note that these two points (either \ding{72} or \ding{108}) do not coincide the maximum location of the two curves (${\cal W}(E_{jet})$), because the two points are obtained by maximizing the {\it total} likelihood Eq. (\ref{eq:KLFitter}), including both transfer function, Gaussian distributions and invariant masses. 

To distinguish the correct and incorrect pairings, we use the ratio of two likelihoods for a given event $x$, $\frac{{\cal L}\left(x|C\right)}{{\cal L}\left(x|W\right)}$. 
Therefore the constant coefficients in Eq. (\ref{eq:KLFitter}) cancel out. 
Requiring $\frac{{\cal L}\left(x|C\right)}{{\cal L}\left(x|W\right)} > 1$, we obtain the 86.6\% purity for parton-level events and 77.6\% purity for detector-level events, keeping 100\% efficiency. 
Although the likelihood analysis is well-motivated, we find in practice that the obtained neutrino momenta are not always sufficiently close to ``true" momenta, which would encourage the use of the matrix element method to be discussed later. It is also difficult to generalize the method, when the masses of particles are unknown.

\subsection{Matrix element method}
\label{sec:mem}

All methods that we discuss in this paper have one thing in common. They all attempt to calculate a good variable for distinguishing different hypotheses. Often these hypotheses are signal plus background and background alone.
The Neyman-Pearson Lemma suggests that the likelihood ratio is the optimal variable to distinguish hypotheses \cite{bishop:2006:PRML,hastie01statisticallearning}.
The likelihood and the probability are the same function with a different choice of dependent and independent variables, so in particle physics, the likelihood could be given by the differential cross section normalized by the total cross section:
\begin{widetext}
\begin{equation}
    {\cal P}(\vec p_i^{\, \rm vis} | \vec \theta \, ) = \frac{1}{\sigma} \int dx_1 dx_2 \frac{f_1(x_1) f_2(x_2)}{2 s x_1 x_2} \times
    \left [ \prod_{j \in {\rm final}} \int \frac{d^3 p_j}{(2\pi)^3 2 E_j} \right ]
    \Big | {\cal M}_{\vec \theta} \, (p_j) \Big |^2
    \prod_{j \in {\rm vis}} {\cal W}(\vec p_j, \vec p_j^{\, \rm vis}) \, , \label{eq:mem}
\end{equation}
\end{widetext}
where $\Big | {\cal M}_{\vec \theta} \, (p_j) \Big |^2$ is the squared matrix element for a given set of parameters $\vec \theta$. This is where the Matrix Element Method (MEM) gets its name. The $x_i$ is the momentum fraction of each parton ($i=1,2$) participating in the collision, $f_i(x_i)$ is the parton distribution function \cite{2015}, and $s$ is the center-of-mass energy of the collider. 
The transfer function ${\cal W}$ parameterizes the detector resolution (as discussed in section \ref{sec:KLfitter}), and the integration is performed over all final state particle momenta (over entire phase space.).  
For the visible final state particles, we integrate over transfer functions. For the invisible final state particles, we integrate over the missing momenta. 

The Matrix Element Method is a type of Multivariate Analysis and provides an optimal variable. However, it can be very challenging to integrate over transfer functions (and accurately parameterizing the detector response in terms of transfer functions) and invisible particle momenta. 
In practice, it may be much easier to get a pretty good  variable by using machine learning techniques on Monte Carlo data. 
Another challenge is how to incorporate the effects of additional radiation and/or other higher order corrections properly \cite{Alwall:2010cq,Campbell:2013hz,Campbell:2013hz}.
Unfortunately the MEM requires the full knowledge of the underlying process including masses, spins and couplings of particles. It is difficult to consider a model-independent analysis. 

Nevertheless, the biggest motivation for using MEM (beyond Neyman-Pearson optimality) is physical transparency.
It is easy to understand where the sensitivity comes from when the discriminating variable is calculated explicitly. 
To get the basic idea on how well MEM could resolve the two-fold ambiguity, let us consider the $gg \to t \bar t \to W^+ W^- b \bar b \to b \bar b \ell^+\ell^- \nu \bar \nu$ process at the parton-level. We reconstruct the two top quarks taking both correct and incorrect combination with {\it true} neutrino momenta. We find that 93.8\% of the time, the squared matrix element for the correct combination is larger. 
This result implies that it would be  difficult to improve beyond 93.8\% using any methods that we are developing. A similar exercise gives 86.2\% for detector-level events. 

For a more realistic investigation, we use \amc~to generate the squared matrix element for $gg \to t \bar t \to W^+ W^- b \bar b \to b \bar b \ell^+\ell^- \nu \bar \nu$. 
By computing the ratio, $\frac{{\cal P}(\vec p^{\, \rm vis} | {\rm  correct})}{{\cal P}(\vec p^{\, \rm vis} | {\rm  incorrect})}$ with MoMEMta (a modular toolkit for the Matrix Element Method at the LHC) \cite{Brochet:2018pqf}, we obtain 84.7\% for the parton-level events and 81.7\% for the detector-level events.

\subsection{Analytic reconstruction}
\label{sec:analytic}

One can in principle solve up to a four-fold ambiguity for the neutrino momentum in the dilepton production using on-shell conditions of the top quark and the $W$-boson \cite{Betchart:2013nba,Sonnenschein:2005ed,Sonnenschein:2006ud,Dalitz:1991wa}. This analytic approach naturally solves the two-fold ambiguity, when trying to reconstruct the final state \cite{Sonnenschein:2005ed,Sonnenschein:2006ud}. However, the method is very sensitive to the exact value of the intermediate particle masses (the top quark and the $W$-boson masses in this case) and the off-shell effects could result in no solution (or imaginary solution). We will not further investigate the features of analytical reconstruction in our current study, since these masses are unknown a priory when applying the method to new physics beyond the SM, and it is difficult to generalize unlike other methods. Finally, we refer to Refs \cite{Matchev:2019bon,Kim:2009si,Park:2021lwa,Rujula:2011qn,Park:2020rol,DeRujula:2012ns} for readers who are interested in the singularity variables.

\bibliography{refs}

\end{document}